\documentclass[aps,pra,twocolumn]{revtex4-1}

\usepackage{graphics}
\usepackage{amsmath}
\usepackage{amssymb}
\usepackage{epsfig}
\usepackage{color}
\usepackage{caption}
\usepackage{subcaption}
\usepackage{bbm}	
\usepackage{adjustbox}
\usepackage{float}
\usepackage{graphicx}   
\usepackage{verbatim}   
\usepackage{hyperref}   

\newcommand* {\Tr}{\ensuremath{\mathrm{Tr}}}
\newcommand{\Real}{\mathbb{R}}

\begin{document}

\title{Positivity violations of the density operator in the Caldeira-Leggett master equation}	

\author{G. Homa} 
\affiliation{  Department of Physics of Complex Systems, E\"{o}tv\"{o}s Lor\'and University, ELTE, P\'azm\'any P\'eter s\'et\'any 1/A, H-1117 Budapest, Hungary}
\email{ggg.maxwell1@gmail.com}
\author{ J. Z. Bern\'ad} 
\affiliation{Department of Physics, University of Malta, Msida MSD 2080, Malta}  
\affiliation{Institut f\"{u}r Angewandte Physik, Technische Universit\"{a}t Darmstadt, D-64289 Darmstadt, Germany}
\email{zsolt.bernad@um.edu.mt} 
\author {L. Lisztes}
\affiliation{Axioplex Ltd., K\"uls\H{o} Szil\'agyi \'ut 36, H-1048, Budapest, Hungary}

	\date{\today}
	
	\begin{abstract}
		{The Caldeira-Leggett master equation as an example of Markovian master equation without Lindblad form is investigated for mathematical consistency. We explore situations both analytically and
			numerically where the positivity violations of the density operator occur. We reinforce some known knowledge about this problem but also find new surprising cases.
			Our analytical results are based on the full solution of the Caldeira-Leggett master equation obtained via the method of characteristics. The preservation of positivity 
			is mainly investigated with the help of the density operator's purity and we give also some numerical results about the violation of the Robertson-Schr\"odinger uncertainty relation.
			}
	\end{abstract}

	\maketitle
	
	\section{Introduction}
	
	States of a quantum mechanical system are given by density operators with spectra consisting only positive eigenvalues and they sum up to one \cite{Neumann}.
	A master equation, governing the time evolution of density operators, has to map density operators to density operators in order to keep the physical interpretation.
	It has been shown by Refs. \cite{GKAS,lindblad1976} that a dynamical evolution given by a semigroup of completely positive maps provides us a Markovian master equation with a generator 
	in Lindblad form. The Markovian master equation obtained within the Caldeira-Leggett model does not belong to the Lindblad class \cite{Ref_3,CL} and thus the generated semigroup is not completely positive. 
	Despite that a stronger condition than positivity fails the dynamical evolution may still map density operators to density operators, a point of view discussed by 
	Refs. \cite{Spohn,Talkner,Ref_4}. However, due to the use of a system plus reservoir model with unitary dynamics and an uncorrelated initial product state 
	the reduced dynamics of the system is always completely positive \cite{Kraus1,book1}. Thus, the completely positive property of the map generated by the Caldeira-Leggett master equation is 
	lost due to the assumptions made in the derivation and proposals to correct this issue have been made since \cite{Ref_1,DIOSI1993517,dekker,Ref_6,Ref_7,Ref_8,Ref_9,Ref_10}.  Without these corrections the 
	Caldeira-Leggett master equation can in principle violate the positivity of the density operator. 
	
	Therefore the circumstances under which the mathematical consistency breaks down are worth investigating. 
	Every now and then, the necessity of clarifying the status of master equations without Lindblad form have initiated investigations in the subject, see for 
	example Ref.\cite{Gnutzmann1996} and the references therein. In fact, the task is to investigate those conditions which forbid the positivity violation of density operator and finally compare 
	them with the approximations used in the derivation of the master equation. Here, we undertake this task with the explicit focus on the Caldeira-Leggett master equation
	where the central system is a harmonic oscillator.
	
	In this paper, we solve exactly the Caldeira-Leggett master equation by using the method of characteristic curves. The idea is based on \cite{RV}, where the differential 
	equation of the density operator in position representation is solved but with an algebraic mistake. It has to be mentioned that in the case of Wigner phase space 
	representations of density operators  the method of characteristic curves has been applied to a more general set of master equations involving also the 
	Caldeira-Leggett master equation \cite{Fleming}. The detailed analysis presented in the latter one does not involve short time evolutions and furthermore only initial
	Gaussian states are considered. Here, we try to identify general conditions for which the Caldeira-Leggett master equation exhibits mathematically inconsistent behavior. Therefore, we have to keep
	track of the whole time evolution from the initial conditions to the steady state. Essentially, we correct and extend the method of Ref. \cite{RV}, which is more adaptable
	in our investigations, due to the technically inconvenient double Fourier transform which connects the Wigner phase space representation with the position representation of the density operator.
	
	The Caldeira-Leggett master equations without the Lindblad form preserves the self-adjointness of the initial density operator. Hence, we require methods which separate
	self-adjoint trace-class operators from density operators. Our first choice is the purity of the density operators, which has to be smaller or equal to one (a necessary but not sufficient condition 
	for a self-adjoint trace-class operator with trace one to be a density operator). A violation of it implies that some of the eigenvalues are not in the interval $[0,1]$, a mathematical inconsistency
	in the physical interpretation of these eigenvalues. The other 
	choice is the Robertson-Schr\"odinger uncertainty relation \cite{SR,Ref_5}, whose derivation is based on the positivity properties of density operators. Thus, deviations from positivity may lead to 
	the violation of the uncertainty relation. The logical implication of these two methods being used to 
	test the positivity of the density operator is very intricate and apart from some comments made in this work we are not
	going to determine it. Thus, we conduct a study on the purity and the Robertson-Schr\"odinger uncertainty relation
	and identify some necessary conditions for the master equation, which guarantee the mathematical consistency.
	
	The paper is organized as follows. In Sec. \ref{I} we discuss some general facts about the purity and the Robertson-Schr\"odinger uncertainty relation. We present the exact solution of the 
	Caldeira-Leggett master equation in Sec. \ref{II}. In Sec. \ref{III} we investigate the conditions for the parameters of the master equation based on steady state solutions. 
	In the next step, the time evolution of the purity and the restrictions to initial conditions are discussed in Sec. \ref{IV}. Numerical
	simulations of both the purity and the Robertson-Schr\"odinger uncertainty relation are collected in Sec. \ref{V}. Technical details, supporting the main text, are given in Appendices \ref{AppI}
	and \ref{AppII}.

	\section{Positivity violation in Markovian master equations}
	\label{I}
	
	A general quantum state is mathematically represented by a self-adjoint positive trace-class 
	operator with unit trace $\hat{\rho}: \mathcal{H}\rightarrow \mathcal{H}$, where $\mathcal{H}$ denotes the complex Hilbert space. 
	A Markovian master equation describing the time evolution of a $\hat{\rho}(t)$ density operator is of the form
	\begin{equation}
	\frac{d\hat{\rho}}{dt}=\mathcal{L} \hat{\rho} \quad \text{with} \quad \Phi_t=e^{ \mathcal{L} t}, \label{evolution}
	\end{equation}   
	and we denote by $\mathcal{D}(\mathcal{H})$ the set of density operators being in the domain of the not necessarily bounded generator $\mathcal{L}$. 
	Mark the set of self-adjoint trace class operators with unit trace $\mathcal{D}^H_1(\mathcal{H})$, being also in the domain of $\mathcal{L}$,  
	and thus implying automatically $\mathcal{D}(\mathcal{H})\subset \mathcal{D}^H_1(\mathcal{H})$. An operator is positive if and only if all of its eigenvalues are greater than or equal to zero, 
	which implies that the eigenvalues of any density operator must satisfy 
	this property. In addition, because the trace of a density operator is one and the trace is just the sum of the eigenvalues, we have that if $\lambda_n$ is an eigenvalue 
	of a density operator, then $0\leq \lambda_n \leq 1$. If
	\begin{equation}
	\exists \left|\Psi\right\rangle:  \quad \left\langle \Psi \left|\hat{O}\right| \Psi\right\rangle <0 \quad \left|\Psi\right\rangle \in \mathcal{H},
	 \nonumber
	\end{equation}
	then this is equivalent to the existence of at least one negative eigenvalue of an operator $\hat{O} \in \mathcal{D}^H_1(\mathcal{H})$.
	
	If we want to determine the set $\mathcal{D}^H_1(\mathcal{H})  \backslash \mathcal{D}(\mathcal{H}) $, then the exact knowledge on the spectrum of all operators
	in $\mathcal{D}^H_1(\mathcal{H})$ is necessary. However, a well-defined subset of $\mathcal{D}^H_1(\mathcal{H})  \backslash \mathcal{D}(\mathcal{H})$ 
	can be filtered out without the complete knowledge of the spectrum by using a simple trick. 
	The next result guarantees the existence of at least one negative eigenvalue.
	
	Let $\hat{\rho}$ be a self-adjoint trace-class operator with 
	$\mathrm{Tr}{\hat{\rho}}=1$. 
	If $\mathrm{Tr}(\hat{\rho}^2)>1$ then $\hat{\rho}$ has at least one negative eigenvalue. The proof of this statement reads as follows. Assume that
	\begin{equation}
	\sum_{j=1}^\infty\lambda_j=1 \quad \text{and} \quad \sum_{j=1}^\infty\lambda_j^2> 1,
	\end{equation}
	where $\lambda_1,\lambda_2,\ldots \in\Real$ denote the eigenvalues of $\hat{\rho}$. Then
	\begin{equation}
	0<\sum_{j=1}^\infty\left(\lambda_j^2-\lambda_j\right),
	\end{equation}
	so there must be a $j_0$ such that $\lambda_{j_0}\not\in[0,1]$; if $\lambda_{j_0}>1$, then there must be a $j_1$ with $\lambda_{j_1}<0$ since $\mathrm{Tr}\hat{\rho}=1$. 
	In the context of the Markovian master equation in \eqref{evolution} we are looking for the class of operators for which the purity  $\mathrm{Tr}(\hat{\rho}^2)$ is larger than one. In this case we say that
	$\hat{\rho} \in \mathcal{D}^P_1(\mathcal{H})\subset\mathcal{D}^H_1(\mathcal{H})  \backslash \mathcal{D}(\mathcal{H})$, because $\mathrm{Tr}(\hat{\rho}^2)\leqslant 1$
	is a necessary but not sufficient condition for $\hat \rho$ to be an element in $\mathcal{D}(\mathcal{H})$. In typical physical models  $\hat{\rho}$ represents a quantum-mechanical system which 
	interacts with an external quantum system and therefore the evolution in \eqref{evolution} is physically consistent if $\hat{\rho}(t) \in \mathcal{D}(\mathcal{H})$ for all $t\geqslant 0$ or being more 
	accurate $\Phi_t$ is a dynamical semigroup with generator $\mathcal{L}$ \cite{Ingarden}. If $\exists t'$ such that 
	$\hat{\rho}(t') \in \mathcal{D}^H_1(\mathcal{H})  \backslash \mathcal{D}(\mathcal{H})$ with initial state $\hat{\rho}_{\text{in}} \in \mathcal{D}(\mathcal{H})$, 
	then the time evolution of the density operator is mathematically inconsistent. If a Markovian master equation generates a uniformly continuous completely positive dynamical semigroup (quantum dynamical semigroup), 
	i.e., $\mathcal{L}$ is bounded and $\mathcal{D}(\mathcal{H})$ is the convex set of all density operators on $\mathcal{H}$,  then it is in Lindblad form \cite{GKAS,lindblad1976,Kraus}
	\begin{gather}
	\frac{d\hat{\rho}}{d t}=\mathcal{L}\hat{\rho}(t)=-\frac{i}{\hbar}\left[\hat{H},\hat{\rho}\right]+\sum_{\alpha}\left[\hat{L}_{\alpha}\hat{\rho}\hat{L}^{\dagger}_{\alpha}-
	\frac{1}{2} \left\{ \hat{L}^{\dagger}_{\alpha} \hat{L}_{\alpha},\hat{\rho}\right\}\right], \nonumber
	\end{gather}
	where $\hat{H}$ is the Hamilton operator and $\hat{L}_{\alpha}$ are the corresponding Lindblad operators. 
	
	There is another possibility to study deviations from the set $\mathcal{D}(\mathcal{H})$.  The Robertson-Schr\"odinger uncertainty relation with 
	$A$ and $B$ essentially self-adjoint operators defined on a dense subset of $\mathcal{H}$ has the following form \cite{SR,Ref_5,Trifonov}
	\begin{equation}
	\sigma_{RS} \geqslant \frac{1}{4}\left|\langle \hat{A}  \hat{B}-\hat{B}  \hat{A} \rangle \right|^2,
	\label{SRI}
	\end{equation}
	with
	\begin{equation}
	\sigma_{RS}=\Delta \hat{A} ^2 \Delta \hat{B} ^2 -\left( \langle  \hat{A}  \hat{B}+\hat{B}  \hat{A} \rangle/2-\langle  \hat{A} \rangle\langle   \hat{B}\rangle\right)^2 , \nonumber
	\end{equation}
	where  $\Delta \hat{\mathcal{O}} ^2=\langle \hat{\mathcal{O}}^2 \rangle- \langle \hat{\mathcal{O}} \rangle^2$ with $\langle \hat{\mathcal{O}} 
	\rangle=\mathrm{Tr} \{ \hat{\mathcal{O}} \hat{\rho}\} $. Since we are going to study
	the Caldeira-Leggett master equation, we will set $\hat A=\hat x$, the position operator, and $\hat B=\hat p$, the momentum operator, which is going to guarantee that the set $\mathcal{D}(\mathcal{H})$ 
	of this specific problem is in the domain of these operators and the right hand side of \eqref{SRI} is simply $\hbar^2/4$ \cite{Ref_5}. The inequality in \eqref{SRI} is based on the 
	positivity of $\hat{\rho}$, which is going to be violated whenever $\hat{\rho} \in \mathcal{D}^I_1(\mathcal{H}) \subset \mathcal{D}^H_1(\mathcal{H})  \backslash \mathcal{D}(\mathcal{H})$. The relation between
	the sets $\mathcal{D}^P_1(\mathcal{H})$ and $\mathcal{D}^I_1(\mathcal{H})$ is not trivial, however we are going to investigate it briefly with the help of numerical simulations. After solving Eq. \ref{evolution} 
	one can determine those time intervals when the solution is in the set  $ \mathcal{D}^H_1(\mathcal{H})  \backslash \mathcal{D}(\mathcal{H}) $ with the help of the purity and/or the  violation of the 
	Robertson-Schr\"odinger uncertainty relation, and of course, if there exists at least one such interval, then the time evolution of the solution is mathematically inconsistent.

	\section{Solution to the Caldeira-Leggett master equation}
	\label{II}
	
	In this section we consider the Caldeira-Leggett master equation \cite{CL} for a quantum harmonic oscillator with  
	frequency $\omega$ ($\hbar=m=k_B=1$)
	\begin{eqnarray}\label{MEQ}
	&&i \frac{\partial \hat{\rho}}{ \partial t}=\left [\frac{\hat{p}^2}{2}+\frac{ \omega^2 \hat{x}^2}{2},\hat{\rho}\right]-i D_{pp}[\hat{x},[\hat{x},\hat{\rho}]]
	+ \gamma [\hat{x},\{\hat{p},\hat{\rho}\}] \nonumber \\
	&&-2 i D_{px}[\hat{x},[\hat{p},\hat{\rho}]],
	\end{eqnarray}
	where $[,]$ stands for commutators while $\{,\}$ for anti-com-\\mutators, and
	$\gamma$ is the relaxation constant. $D_{pp}$ is the momentum diffusion coefficient and finally $D_{px}$ is the cross diffusion coefficient. This master equation is derived from the 
	Caldeira-Leggett model \cite{CL}, where an environment of harmonic oscillators in thermal equilibrium with temperature $T$ is considered with Ohmic spectral density and a high 
	frequency cut-off $\Omega$. The central system, a harmonic oscillator in our case, is taken to be slow compared to the bath correlation time $\Omega, T \gg \omega$ and the Born-Markov approximation
	$\Omega, T \gg \gamma$ is also employed during the derivation. The master equation in (\ref{MEQ}) is not in Lindblad form and therefore, according to the introductory notions in Sec. \ref{I} there may be 
	time intervals where $\hat{\rho}(t) \in \mathcal{D}^H_1(\mathcal{H})  \backslash \mathcal{D}(\mathcal{H})$. In order to examine closely these situations we are going to determine the exact solution of this master 
	equation.
	
	As a first step we rewrite Eq. \ref{MEQ} in the position representation
	\begin{eqnarray}
	&&i \frac{\partial}{\partial t} \rho(x,y,t)=\Big[ \frac{1 }{2}\left(\frac{\partial^2}{\partial y^2}-
	\frac{\partial^2}{\partial x^2} \right) +\frac{ \omega^2}{2} \left(x^2-y^2\right) \nonumber \\
	&&-i D_{pp} (x-y)^2 
	-i \gamma (x-y) \left(\frac{\partial}{\partial x}-
	\frac{\partial}{\partial y} \right) \nonumber \\
	&&-2 D_{px} (x-y) \left(\frac{\partial}{\partial x}+
	\frac{\partial}{\partial y} \right) \Big]  \rho(x,y,t). \nonumber
	\end{eqnarray}
	We introduce the center of mass and relative coordinates $R=(x+y)/2$, $r=x-y$, and the master equation becomes 
	\begin{eqnarray}
	&& \frac{\partial \rho(R,r,t)}{\partial t}=\Big[i\frac{\partial^2}{\partial r \partial R} - 2 \gamma r \frac{\partial}{\partial r}- D_{pp} r^2 + 2 i D_{px}r\frac{\partial}{\partial R} \nonumber \\ && + 
	\frac{\omega^2 r R}{i}\Big]\rho(R,r,t). \nonumber
	\end{eqnarray}
	Fourier transforming the equation in the variable $R$ reduces it to a first order partial differential equation 
	\begin{eqnarray}
	&&\left[ \left( 2 \gamma r - K\right) \frac{\partial}{\partial r}+ \omega^2 r\frac{\partial}{\partial K}+ D_{pp} r^2 - 2 D_{px}r K \right]\rho(K,r,t) \nonumber \\ 
	&& =-\frac{\partial \rho(K,r,t)}{\partial t}, \nonumber
	\end{eqnarray}
	where we used the following identity
	\begin{equation}
	\rho(R,r,t)= \frac{1}{\sqrt{2 \pi}} \int_{-\infty}^{\infty}{dK \exp{(i KR)} \rho(K,r,t)}. \nonumber
	\end{equation}
	
	This equation was written by Roy and Venugopalan in \cite{RV} for the case of  $D_{px}=0$, but we found that their solution did not satisfy the master equation for every instance of 
	time due to an algebraic mistake. They used the method of characteristics to find the solution of this equation, and we applied the same technique for the extended case of 
	$D_{px} \neq 0$.
	
	The method of characteristics (see \cite{MOC}) is a very useful technique for solving first order partial differential equations by reducing  partial differential equations to a family of ordinary 
	differential equations along which the solution can be integrated from some initial data given on a suitable hypersurface. In our case the method of characteristics leads to the following system of 
	ordinary differential equations:
	\begin{eqnarray}
	&&\frac{\partial}{\partial t}r \left( t \right) = 2
	\gamma\,r \left( t
	\right) -K \left( t \right) \nonumber \\
	&&  \frac {\partial}{\partial t}K \left( t \right) = {\omega}^{2}r \left( t
	\right) \nonumber \\
	&&\frac {\partial}{\partial t}\rho \left( t \right) = \left(\, 2 D_{px}r\left(t\right) K\left(t\right) - D_{pp}  r^2 \left( t \right)\right) \rho \left( t
	\right) \nonumber
	\end{eqnarray}
	with initial conditions $r(0)=r_0, K(0)=K_0, \rho(K_0,r_0,0)=\rho_0$. In order to obtain simple formulas we introduce the following notations 
	\begin{eqnarray}
	&& Y:= {\omega}^{2}r_{0}-K_0 \lambda_1, \quad \Gamma:=r_0 \lambda_2-K_{0}, \quad \lambda_{1,2}:=\gamma \pm \sqrt{X}, \nonumber \\
	&& X:=\gamma^2-\omega^2 \nonumber
	\end{eqnarray}
	and with these markings, the solution of the ordinary differential equation system is as follows
	\begin{align}
	K \left( t \right): =\frac {Y\,{ e^{\lambda _{1}\,t}}-{e^{\lambda _{2}\,t}
		} \left( Y-2 K_{0}\sqrt{X}
		\right) }{2\sqrt{X}}, \nonumber
	\end{align}
	and
	\begin{equation}
	r(t):=\frac {\Gamma\,{e^{\lambda _{1}\,t}}-{e^{\lambda _{2}\,t}
		} \left( \Gamma-2r_{0}\sqrt{X}
		\right) }{2\sqrt{X}}. \nonumber
	\end{equation}
	A compact form of the final results reads
	\begin{equation}
	\rho(t)=\rho _{0} \exp\left[-\frac{A e^{-2\,t \left( 2\, \sqrt{X}+\gamma
			\right) }D_{pp}+B e^{-2\,t \left( 2\, \sqrt{X}+\gamma
			\right) }D_{px}}{8{X}^{3/2}\gamma \omega^2}\right], \nonumber
	\end{equation}
	where the notations $A$ and $B$ can be found in Appendix \ref{AppI}. After substituting $K(t)\rightarrow K, r(t)\rightarrow r$, and $\rho(t)\rightarrow \rho(K,r,t)$,  we get
	\begin{eqnarray}
	&&K_0=\frac {   \left[  \left(\sqrt{X}+\gamma \right) K-
		\omega^{2}r \right] e^{2\sqrt{X}t}+ \left(\sqrt{X}-\gamma\right) K+\omega^{2}r}{
		2 \sqrt{X} e^{\lambda_{1}t}},  \nonumber \\
	&&r_0=\frac { \left[\left(\sqrt{X}-\gamma \right)r+K \right]
		e^{2 \sqrt{X}t}+ \left(\sqrt{X}+\gamma \right)r-K}{2 \sqrt{X} e^{\lambda _{1}t}}. \nonumber
	\end{eqnarray}
	Thus,
	\begin{equation}
	\rho_0=\rho (K,r,t) e^{\frac {A e^{-2\,t \left( 2\, \sqrt{X}+\gamma
				\right) }D_{pp}+B e^{-2\,t \left( 2\, \sqrt{X}+\gamma
				\right) }D_{px}}{8{X}^{3/2}\gamma \omega^2}}, \nonumber
	\end{equation}
	and finally the exact solution at an arbitrary time is
	\begin{equation}
	\rho (K,r,t)=\rho_0 e^{-\frac {Ae^{-2\,t \left( 2\, \sqrt{X}+\gamma
				\right) }D_{pp}+B e^{-2\,t \left( 2\, \sqrt{X}+\gamma
				\right) }D_{px}}{8{X}^{3/2}\gamma \omega^2}}. 
	\label{solution}
	\end{equation}
	
	\section{Some requirements for the parameters of the master equation}
	\label{III}
	
	In this section we analyze some relations between the parameters of the master equations. We reintroduce for the physically important relations the dimensions of all
	parameters ($\hbar,m, k_B \neq 1$) and for the pure mathematical observations we avoid them again in order to obtain simple formulas. Let us consider the steady state of the master equation 
	\eqref{MEQ} 
	in position representation with reintroduced physical dimensions 
	\begin{eqnarray}
	&&\rho(x,y,\infty)=\frac {\sqrt{\gamma}m\omega}{\hbar \sqrt {\pi }\sqrt {D_{pp}-4 \gamma m D_{px}}} \times \nonumber \\ &&\exp\left\{-\frac{\gamma \left[  m \omega \left( x+y \right)\right]^2}
	{4 \hbar^2 \left(D_{pp}-4 \gamma m D_{px}\right)}-\frac{D_{pp}\left(x-y \right)^2}{4 \gamma} \right\}, \nonumber
	\end{eqnarray}
	which agrees with the steady state known in the literature for $D_{px}=0$, see for example \cite{book1}. The eigenproblem of the steady state reads \cite{BCSH}
	\begin{equation}\label{rhoeigen}
	\int_{-\infty}^{+\infty}\rho(x,y, \infty )\phi_n(y) \,dy=\epsilon_n\phi_n(x) \nonumber
	\end{equation}
	and is solved by
	\begin{eqnarray}
	&&\phi_n(x)=H_n\left(x,\frac{1}{4 \sqrt{AC}}\right)\exp\left\{-2\sqrt{AC} x^2 \right\},\\
	&&\epsilon_n=\epsilon_0\epsilon^n, \quad \epsilon_0=\frac{2\sqrt{C}}{\sqrt{A}+\sqrt{C}}, \quad \epsilon=\frac{\sqrt{A}-\sqrt{C}}{\sqrt{A}+\sqrt{C}}, \nonumber \\
	&&A=\frac{D_{pp}}{4\gamma}, \quad C=\frac{\gamma \left(  m \omega \right)^2}
	{4 \hbar^2 \left(D_{pp}-4 \gamma m D_{px}\right)}, \nonumber
	\end{eqnarray}
	where $H_n(x,a)$ is a generalized Hermite polynomial with $a>0$. Therefore, the model is not physical if the following inequality does not hold: $A \geqslant C$ or
	\begin{equation}
	\label{general}
	\hbar^2 \frac{D_{pp}^2-4 \gamma m D_{pp} D_{px}}{\gamma^2 m^2 \omega^2}\geqslant 1. 
	\end{equation}
	If this inequality is not satisfied  then the stationary density operator has negative eigenvalues for every odd  $n \in \mathbb{N}$. 
	
	\begin{table} \scriptsize
		\centering
		\begin{adjustbox}{width=0.47\textwidth}
			\begin{tabular}{||c||c||c||}
				\hline
				\hline 
				Case  &   $D_{pp}$ & $D_{px}$  \\
				\hline
				I \cite{CL}  &  $2 \gamma m k_B T/\hbar^2$  &  $0$    \\
				\hline
				II \cite{book1}   & $2 \gamma m k_B T/\hbar^2$  &  $-\gamma k_B T/ \left(\hbar^2 \Omega\right)$  \\
				\hline
				III \cite{DIOSI1993517}  &  $2 \gamma m k_B T/\hbar^2$   & $ \Omega \gamma/ \left(6 \pi k_B T \right)$ \\
				\hline
				IV \cite{dekker}  &  $\gamma m \omega^2/\left(2 \hbar \sqrt{\omega^2-\gamma^2} \right)$ & $\gamma^2/\left(\hbar  \sqrt{\omega^2-\gamma^2} \right) $   \\
				\hline
				\hline
			\end{tabular} 
		\end{adjustbox}
		\caption{Diffusion's coefficients in literature}  \label{tab:table1}
	\end{table}
	
	The coefficients $D_{pp}$ and $D_{px}$ of the Caldeira-Leggett master equation in \eqref{MEQ} are determined in the high temperature limit $k_B T \geqslant \hbar \Omega \gg \hbar \omega$. 
	First, we consider the master equation of Ref. \cite{CL}, see case I in Table \ref{tab:table1}. Substituting into Eq. \eqref{general} we obtain
	\begin{equation}
	T^{\text{I}}_{\text{min}} \geqslant \frac{1}{2} \frac{\hbar \omega}{k_B}, \label{Tmin1}
	\end{equation}
	a minimum temperature for the environment, which guarantees that the steady state has no negative eigenvalues. This condition is in accordance with the high temperature limit 
	$k_B T \gg \hbar \omega$ employed
	in the derivation of the master equation, thus is always fulfilled. Secondly, one may consider the Markovian limit of the more general non-Markovian master equation of the Caldeira-Leggett
	model also in the high temperature limit, see case II in Table \ref{tab:table1} \cite{book1,KRGH,HFRR,Hu_Paz}. Now, Eq. \eqref{general} yields
	\begin{equation}
	T^{\text{II}}_{\text{min}} \geqslant \frac{1}{2} \frac{\hbar \omega}{k_B \sqrt{1+2 \gamma/\Omega}}, \label{Tmin2}
	\end{equation}
	which is always fulfilled due to the Born-Markov approximation $\Omega \gg \gamma$ and the high temperature limit $k_B T \gg \hbar \omega$.
	
	In the following, we consider two Markovian master equations which are in Lindblad form \cite{DIOSI1993517,dekker}, but neglecting the position diffusion term $-D_{xx} [\hat{p},[\hat{p},\hat{\rho}]]$, we arrive at 
	the same evolution as in Eq. \eqref{MEQ}. This brief study is motivated by the fact that in experiments the position diffusion has not been detected yet. Therefore, it is worth to ask what conditions do the 
	parameters of the truncated master equations have to fulfill. In the case of \cite{DIOSI1993517}, where medium temperatures are considered 
	$\hbar \Omega\geqslant k_B T \gg \hbar \omega$, Eq. \eqref{general} yields
	\begin{equation}
	T^{\text{III}}_{\text{min}} \geqslant \frac{1}{2} \frac{\hbar \omega}{k_B} \sqrt{1+\frac{2}{3\pi}\frac{\Omega \gamma}{\omega^2}}, \label{Tmin3}
	\end{equation}
	see the coefficients $D_{pp}$ and $D_{px}$ in the case III of Table \ref{tab:table1}. This condition is true only when the cutoff energy $\hbar \Omega$ is not too large
	compared with $k_B T$. The second truncated Markovian master equation of Ref. \cite{dekker} is obtained via a phenomenological phase space quantization of an underdamped $\omega >\gamma$ harmonic
	oscillator. Thus no model for the environment is required, i.e., $T=\Omega=0$, yielding the coefficients $D_{pp}$ and $D_{px}$ in case IV of Table \ref{tab:table1}. Hence, for \eqref{general} we have
	\begin{equation}
	0\geqslant 3 \omega^2 + 4 \gamma^2, \nonumber
	\end{equation}
	which is true only if $\omega=\gamma=0$. Therefore, the master equation of \cite{dekker} cannot be truncated at will. 
	
	In the subsequent discussion, we focus only on the first three cases of Table \ref{tab:table1} and investigate the behavior of $\Tr{\hat{\rho}^2}$ as a function of temperature $T$. As it will be a 
	purely mathematical discussion we return to the convention $\hbar=k_B=m=1$. Applying the Plancherel theorem \cite{yosida} we 
	can get the purity in the following form:
	\begin{equation}
	\Tr(\hat{\rho}^2)=\int_{-\infty}^{\infty}{\int_{-\infty}^{\infty}{\rho(K,r,t)\rho^{\ast}(K,r,t) dK dr}}. \nonumber 
	\end{equation}
	Taking into account (\ref{solution}) the purity becomes
	\begin{eqnarray}
	&& \Tr(\hat{\rho}^2)=\int_{-\infty}^{\infty}\int_{-\infty}^{\infty} |\rho(K_0,r_0,0)|^2 \times \nonumber \\  &&\left|\,{e^{-\,{\frac {A{e^{-2\,t \left( 2\, \sqrt{X}+\gamma
							\right) }}D_{pp}+B{e^{-2\,t \left( 2\, \sqrt{X}+\gamma
							\right) }}D_{px}}{8{X}^{3/2}\gamma\,{\omega}^{2}}}}}\right|^2 dK dr, \label{Plan} 
	\end{eqnarray}
	where the details about $A$ and $B$ can be found in Appendix \ref{AppI} and $X=\gamma^2-\omega^2$. Furthermore, both
	$D_{pp}$ and $D_{px}$ are functions of $T$. In the following we will show that the  derivative of the purity with respect to 
	temperature $\partial \Tr(\hat{\rho}^2) / \partial T$, regardless of the  initial conditions and other parameters,
	\begin{eqnarray}
	&& \frac{\partial \Tr(\hat{\rho}^2) }{\partial T}=2 \Re \int_{-\infty}^{\infty}\int_{-\infty}^{\infty} \left|\rho(K_0,r_0,0)\right|^2 \times \nonumber \\ 
	&& \,{e^{2 \Re\left[-\,{\frac {A{e^{-2\,t \left( 2\, \sqrt{X}+\gamma
							\right) }}D_{pp}+B{e^{-2\,t \left( 2\, \sqrt{X}+\gamma
							\right) }}D_{px}}{8{X}^{3/2}\gamma\,{\omega}^{2}}}\right]}}  \times \nonumber  \\ 
	&& \frac{\partial}{\partial T } \left(-\,{\frac {A{e^{-2\,t \left( 2\, \sqrt{X}+\gamma
					\right) }}D_{pp}+B{e^{-2\,t \left( 2\, \sqrt{X}+\gamma
					\right) }}D_{px}}{8{X}^{3/2}\gamma\,{\omega}^{2}}}\right)  dK dr \nonumber
	\end{eqnarray}
	is always non-positive.  $\Re$  denotes the real part of a complex number. 
	
	Since case I is just a simplified version of case II in Table \ref{tab:table1} with $D_{px}=0$, we are going to consider case II in the subsequent discussion. 
	Let us introduce function $H(K,r,t)$ as follows
	\begin{eqnarray}
	&&H(K,r,t):=G(K,r,t)+c = \nonumber \\
	&&=2 \Re\ \frac{\partial}{\partial T } \left(-\,{\frac {A{e^{-2\,t \left( 2\, \sqrt{X}+\gamma
					\right) }}D_{pp}+B{e^{-2\,t \left( 2\, \sqrt{X}+\gamma
					\right) }}D_{px}}{8{X}^{3/2}\gamma\,{\omega}^{2}}}\right), \nonumber
	\end{eqnarray}
	where $G(K,r,t)$ is given in Appendix \ref{AppI} and
	\begin{equation}
	c=-\frac{1}{2}\,{\frac { \left( -2\,{K}^{2}\gamma+\Omega\, \left( {\omega}^{2}{r
			}^{2}+{K}^{2} \right)  \right) T}{\Omega\,{\omega}^{2}}} \leq 0. \nonumber
	\end{equation}
	The maximum of $H(K,r,t)$ is at $t=0$, because it is exponentially decreasing with time, hence it is enough to investigate the  derivative of the purity with respect to temperature at $t=0$. 
	After a brief calculation we get:
	\begin{equation}
	H(K,r,0)=0, \nonumber
	\end{equation}
	and consequently
	\begin{eqnarray}
	&&\frac{\partial \Tr(\hat{\rho}^2) }{\partial T}=\int_{-\infty}^{\infty} dK\int_{-\infty}^{\infty} dr \left|\rho(K_0,r_0,0)\right|^2 \times \nonumber \\ 
	&&e^{2 \Re\left[-\,{\frac {A{e^{-2\,t \left( 2\, \sqrt{X}+\gamma
						\right) }}D_{pp}+B{e^{-2\,t \left( 2\, \sqrt{X}+\gamma
						\right) }}D_{px}}{8{X}^{3/2}\gamma\,{\omega}^{2}}}\right]}H(K,r,t)   \nonumber \\
	&&\leq \int_{-\infty}^{\infty} dK\int_{-\infty}^{\infty} dr \left|\rho(K_0,r_0,0)\right|^2 \times \nonumber \\ &&e^{2 \Re\left[-\,{\frac {A{e^{-2\,t \left( 2\, \sqrt{X}+\gamma
						\right) }}D_{pp}+B{e^{-2\,t \left( 2\, \sqrt{X}+\gamma
						\right) }}D_{px}}{8{X}^{3/2}\gamma\,{\omega}^{2}}}\right]}H(K,r,0)  \nonumber \\
	= 0 \nonumber
	\end{eqnarray}
	for all $t$ and all possible parameter values and initial conditions.
	
	Taking case III from Table \ref{tab:table1} $H(K,r,t)$ reads now as follows
	\begin{eqnarray}
	&& H(K,r,t):=F(K,r,t)+k(K,r)= \nonumber \\
	&& 2 \Re\ \frac{\partial}{\partial T } \left(-\,{\frac {A{e^{-2\,t \left( 2\, \sqrt{X}+\gamma
					\right) }}D_{pp}+B{e^{-2\,t \left( 2\, \sqrt{X}+\gamma
					\right) }}D_{px}}{8{X}^{3/2}\gamma\,{\omega}^{2}}}\right), \nonumber
	\end{eqnarray}
	where
	\begin{equation}
	k(K,r)=-{\frac {3\,{T}^{2} \left( {\omega}^{2}{r}^{2}+{K}^{2} \right) 
			\pi +{K}^{2}\Omega\,\gamma}{3\pi \,{T}^{2}{\omega}^{2}}} \nonumber
	\end{equation}
	and details about $F(K,r,t)$ are shown in Appendix \ref{AppI}. Starting from this point the course of the proof is the same as in case II and ultimately the derivative of the purity with respect to temperature 
	$\partial \Tr(\hat{\rho}^2) / \partial T$, regardless of the initial conditions and other parameters is always non-positive in this case as well. Thus, whenever 
	$\hat{\rho} \in \mathcal{D}^P_1(\mathcal{H})$ or $\mathrm{Tr}(\hat{\rho}^2)>1$, by the increase of the temperature the purity is decreased and the positivity violation of the 
	density operator might be corrected for a given temperature provided that the condition $\mathrm{Tr}(\hat{\rho}^2)\leqslant1$ is not fulfilled in a pathological way. Here, we remind the reader
	about the condition $\mathrm{Tr}(\hat{\rho}^2)\leqslant1$, which is necessary but not sufficient for $\hat{\rho}$ to be a density operator. 
	
	In summary, we made a brief analysis on the steady state of \eqref{MEQ} and we found in regard to three Markovian master equations (cases I, II, III in  Table \ref{tab:table1}) known in the 
	literature that the steady state does not violate the positivity for the parameter ranges used in the approximations of the derivations. In the case IV of Ref. \cite{dekker} the master equation 
	cannot be truncated to obtain \eqref{MEQ}, because the steady state will be no longer a density operator. Furthermore, the purities of the density operators in cases I, II, III are
	monotonically decreasing with temperature, which means that the study of purity for the positivity violation becomes obsolete for very high temperatures. Therefore, we turn our attention 
	to the temporal behavior of $\Tr\hat{\rho}^2$ and keeping the temperature within the range of approximations of the master equations, but not too high.
	
	\section{Temporal behavior of the purity and physical conditions for pure initial states}
	\label{IV}
	
	In this section we investigate the temporal behavior of the purity for all cases in Table \ref{tab:table1}. The main issue is in fact the situation when the purity is one at a particular time
	and its derivative with respect to time is positive, thus the purity will increase above one for later times. A special case of this situation is when we have a pure initial state. 
	
	The time evolution of the purity is governed by the following differential equation
	\begin{eqnarray}
	\frac{1}{2} \frac{\partial \Tr \hat{\rho}^2}{\partial t}&=& \gamma\Tr \hat{\rho}^2+ 2 D_{pp} \Tr\left[\left(\hat{x}\hat{\rho}\right)^2-\hat{x}^2\hat{\rho}^2\right]+4 D_{px} \nonumber \\
	&\times& \Tr\left[\left(\hat{x}\hat{\rho}\right)\left(\hat{p}\hat{\rho}\right)-\left(\frac{1}{2}\hat{I}+\hat{p}\hat{x}\right)\hat{\rho}^2\right]  \nonumber \\
	&=&
	\left(\gamma-2 D_{px}\right)\Tr \hat{\rho}^2 +4 D_{px} F_1(t)+\frac{1}{2}F_2(t),\nonumber 
	\end{eqnarray}
	where we have used equation \eqref{MEQ} and introduced the notations:
	\begin{eqnarray}
	F_1(t)&=&\Tr\left[\left(\hat{x}\hat{\rho}\right)\left(\hat{p}\hat{\rho}\right)-\hat{p}\hat{x}\hat{\rho}^2\right] \nonumber \\
	F_2(t)&=&4 D_{pp} \Tr\left[\left(\hat{x}\hat{\rho}\right)^2-\hat{x}^2\hat{\rho}^2\right]. \nonumber
	\end{eqnarray}
	For the sake of simplicity let us introduce $P(t)=\Tr \hat{\rho}^2$, which yields
	\begin{equation}
	\frac {\partial}{\partial t} P \left( t \right) = \left( 2\,\gamma-4\,D_{px} \right) P \left( t \right) +8\,D_{px}\,F_1(t)+F_2(t) \nonumber
	\end{equation}
	and the solution to this differential equation with initial condition $P(0)$ is
	\begin{eqnarray}
	&&P \left( t \right) =P(0) \, e^{2\,( \gamma-2\,D_{px}) t} \nonumber \\
	&&  +\int_{0}^{t} e^{2\,( -\gamma+2\,D_{px}) (t'-t)} \left[ 8\,D_{px}\,F_1(t')+F_2(t')  \right] \,{\rm d}t'.  \nonumber
	\end{eqnarray}
	
	It is interesting to note the case when $\hat{x}\hat{\rho}$ is a Hilbert-Schmidt operator \cite{yosida}. As the Hilbert-Schmidt operators
	form a Hilbert space, the Cauchy-Schwartz-Bunyakovsky inequality of the inner product for $\hat{x}\hat{\rho}$ yields
	\begin{eqnarray}
	&&\Tr\left[(\hat{\rho}\hat{x})^{\dagger}(\hat{x}\hat{\rho})\right] \leq \sqrt{\Tr\left[(\hat{\rho}\hat{x})(\hat{\rho}\hat{x})^{\dagger}\right]}\sqrt{\Tr\left[(\hat{x}\hat{\rho})(\hat{x}
		\hat{\rho})^{\dagger}\right]} \nonumber \\
	&&=\sqrt{\Tr\left[(\hat{\rho}^2\hat{x}^2)\right]}\sqrt{\Tr\left[\hat{x}^2\hat{\rho}^2\right]}=\Tr\left[\hat{\rho}^2\hat{x}^2\right], \nonumber
	\end{eqnarray}
	which results $F_2(t)\leq0$. 
	
	The purity $P(t)$ is monotonically decreasing or constant at a time $t$ if 
	\begin{equation}
	\left( 2\,\gamma-4\,D_{px} \right) P \left( t \right) +8\,D_{px}\,F_1(t)+F_2(t) \leqslant 0. \nonumber
	\end{equation}
	This inequality must be fulfilled at any time $t$ when the purity $P(t)=1$, i.e., $\hat{\rho}$ is a pure state. Hence, we have 
	\begin{eqnarray}
	1-2F_1(t)&=&\langle \hat{x}\hat{p}+\hat{p}\hat{x} \rangle-2\langle \hat{x}\rangle \langle \hat{p}\rangle=2\sigma^2_{px}, \nonumber \\
	F_2(t)&=&-4 D_{pp} \left( \langle \hat{x}^2\rangle-\langle \hat{x} \rangle^2 \right)= -4 D_{pp} \sigma^2_{xx}, \nonumber
	\end{eqnarray}
and
	\begin{equation}
	\gamma-2\left(D_{pp} \sigma^2_{xx}+2 D_{px}  \sigma^2_{px}\right) \leqslant 0. \label{cond}
	\end{equation}
	
	In general, the above condition applies to all situations when the state is pure. However, the most convenient way of application is for 
	initial pure states. Therefore, we consider an initial pure state and all parameters have dimensions ($\hbar,m, k_B \neq 1$). Case I from Table \ref{tab:table1} results in
	\begin{equation}
	\frac{1}{2} \frac {\hbar}{\sqrt{m k_B T}}\leq \sigma_{xx}, \nonumber
	\end{equation}
	which means that $\sigma_{xx}$--the width of the initial wave packet-- has to be approximately five times larger than the thermal wavelength.  A condition,
	which is not satisfied by many initial states, but for example all the eigenstates of the quantum harmonic oscillator are subject to \eqref{cond} due to
	the high temperature limit $k_B T \gg \hbar \omega$ approximation employed in derivation of the master equation of case I.
	
	In case II we have
	\begin{equation}
	\frac{1}{2} \sqrt{\frac{\hbar^2}{m k_B T}+\frac{2 \sigma^2_{px}}{m \Omega}} \leqslant \sigma_{xx}, \nonumber
	\end{equation}
	which is a very similar condition to case I and it is satisfied also by all the eigenstates of the quantum harmonic oscillator. 
	
	Considering the truncated master equations of Refs. \cite{DIOSI1993517,dekker}, we are able to obtain conditions for the initial states. Case III from Table \ref{tab:table1}  
	yields 
	\begin{equation}
	\sqrt{\frac{\hbar^2}{4 m k_B T}-\frac{\hbar^2 \Omega}{12 \pi m \left(k_B T\right)^2} \sigma^2_{px}}\leqslant \sigma_{xx}, \nonumber
	\end{equation}
	and the self-consistency condition
	\begin{equation}
	\sqrt{\frac{3 \pi k_B T}{ \Omega}} \geqslant \sigma_{px}. \nonumber
	\end{equation}
	Surprisingly, we get the same conclusion obtained in the investigation of the steady state (see Sec. \ref{III}), namely the cutoff energy $\hbar \Omega$ cannot
	be too large compared with $k_B T$. Otherwise the self-consistency condition reads: $\sigma_{px}$ is smaller or equal to a very small number, which most of
	the pure states do not satisfy.
	
	Finally, case IV from Table \ref{tab:table1} results in
	\begin{equation}
	\sqrt{\frac{\hbar \sqrt{\omega^2-\gamma^2}}{m \omega^2}-\frac{2 \gamma}{m \omega^2}\sigma^2_{px}} \leqslant \sigma_{xx}, \nonumber
	\end{equation}
	and the self-consistency condition
	\begin{equation}
	\sqrt{\frac{\hbar  \sqrt{\omega^2-\gamma^2}}{2\gamma}} \geqslant \sigma_{px}. \nonumber
	\end{equation}
	These conditions can be subject to many initial pure states, but we already know from Sec. \ref{III} that the truncated master equations of case IV violates
	positivity of the steady state. Thus, the truncated master equation of Ref. \cite{dekker} may be applied for short time evolutions but definitely not for longer times. 
	
	In this section we obtained conditions for initial pure states by studying the time evolution of the purity. Thus, those initial pure states, which fulfill these conditions,
	guarantee a short time evolution where the value of the purity does not exceed one. The Caldeira-Leggett master equation, which is covered by cases I and II, shows for example that also the
	eigenstates of the quantum harmonic oscillator are subject to these conditions. One may think, with these type of initial pure states the purity never exceeds one. In contrary this can happen 
	and we will present our numerical experiences in the next section.
	
	\begin{figure*}[ht!]
		\begin{subfigure}{.49\textwidth}
			\includegraphics[trim={8.5cm 20cm 0 20cm},clip,width=12cm]{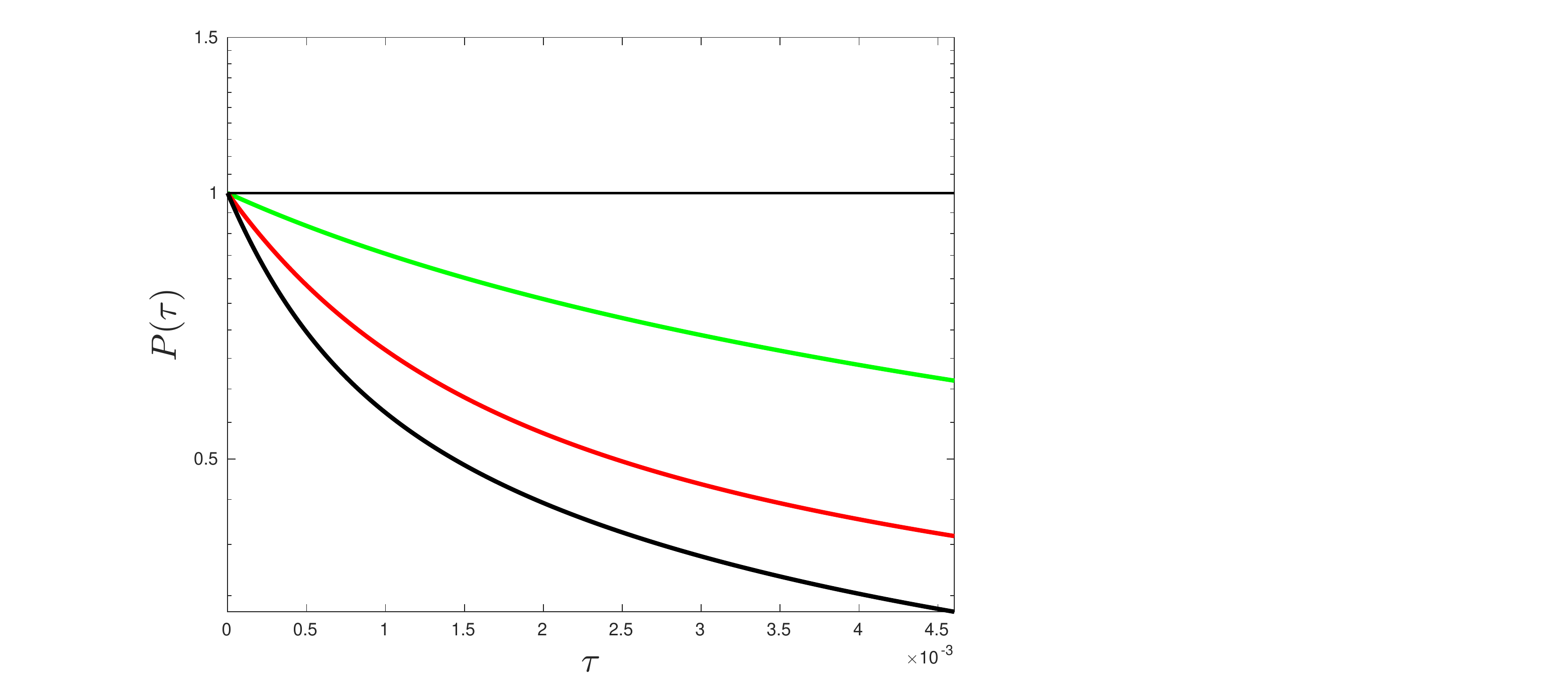}
			\caption{}
		\end{subfigure}
		\begin{subfigure}{.49\textwidth}
			\includegraphics[trim={8.5cm 20cm 0 20cm},clip,width=12cm]{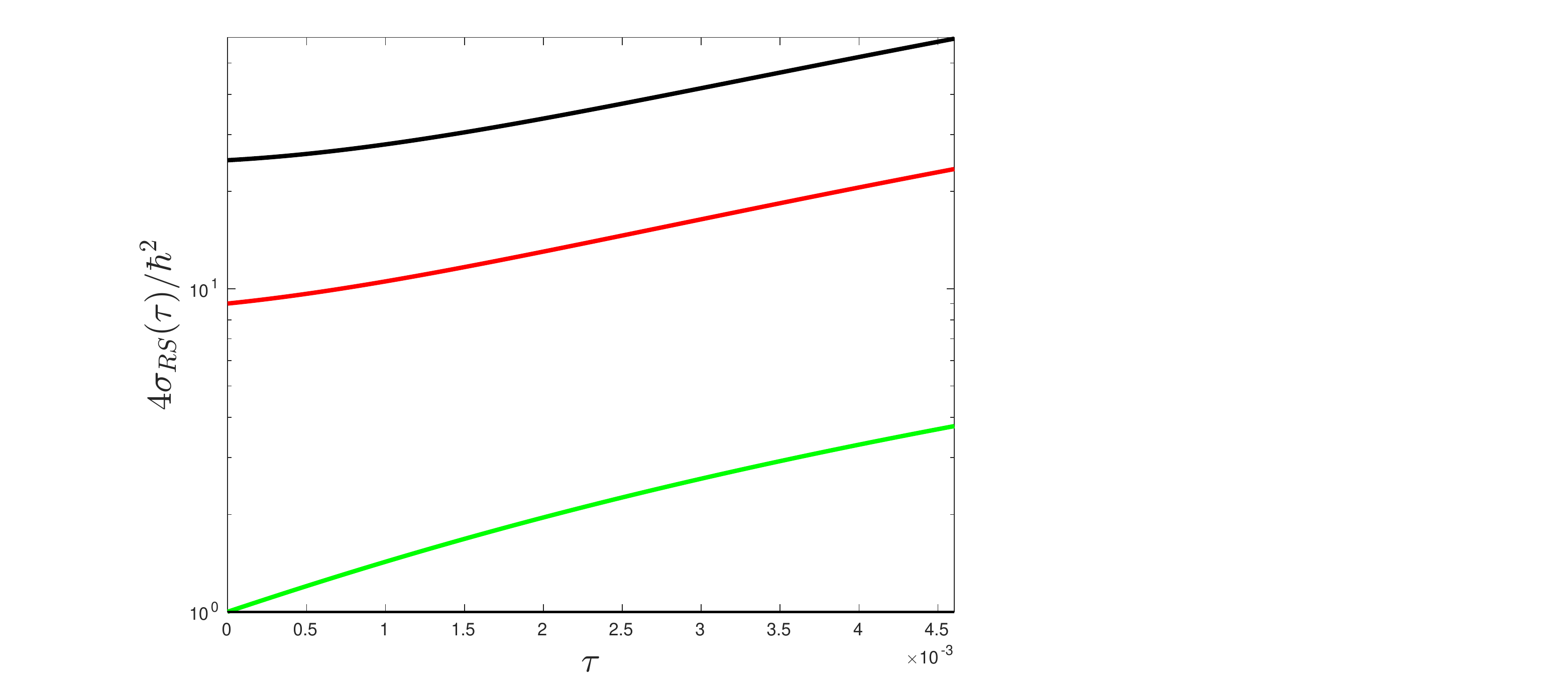}
			\caption{}
		\end{subfigure}
		\caption{Left panel: Semilogarithmic plot of purity $P(\tau)$ as a function of dimensionless time $\tau=\omega t$. Right panel: Left hand-side of the Robertson-Schr\"odinger uncertainty relation in \eqref{SRI}, used in dimensionless 
			form $4\sigma_{RS}(\tau)/\hbar^2$, as a function of $\tau$. The three different curves are plotted for the following initial states of Eq. \eqref{hos}: $n=0$ (green); $n=1$ (red); and $n=2$ (black). 
			We set $D_{px}=0$, $\gamma'=0.35$, $T'= T'^{I}_{min}$ (see Eq. \eqref{Tmin1}), and $\beta'=0.6$ according to Eq. \eqref{transf}. 
			The solid black lines mark both the allowed upper bound for the purity in the left panel and the right hand-side of the Robertson-Schr\"odinger uncertainty relation in the right panel. In the short time evolution, 
			the purity does not exceed one and the Robertson-Schr\"odinger uncertainty relation is not violated.} \label{fig1}
	\end{figure*}
	
	\begin{figure*}[ht!]
		\begin{subfigure}{.49\textwidth}
			\includegraphics[trim={8.5cm 20cm 0 20cm},clip,width=12cm]{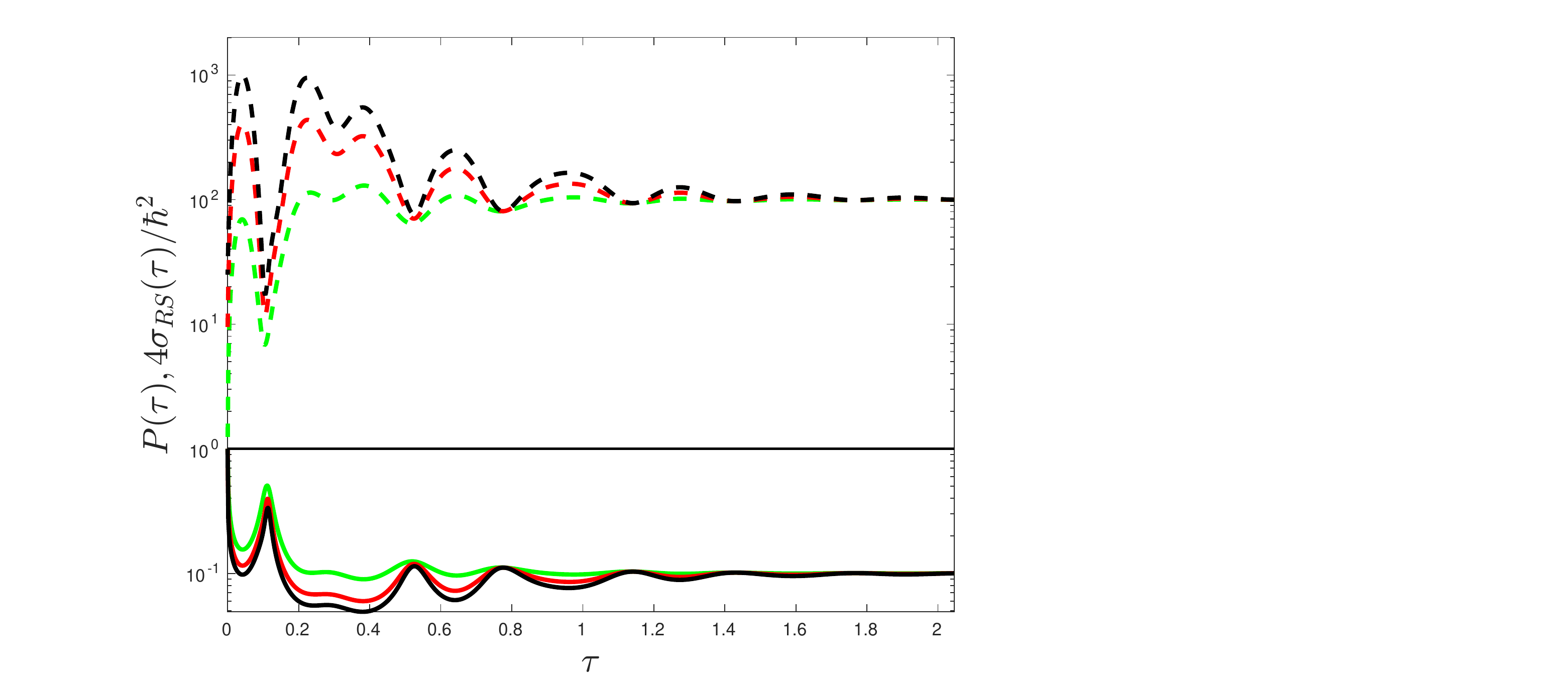}
			\caption{}
		\end{subfigure}
		\begin{subfigure}{.49\textwidth}
			\includegraphics[trim={8.5cm 20cm 0 20cm},clip,width=12cm]{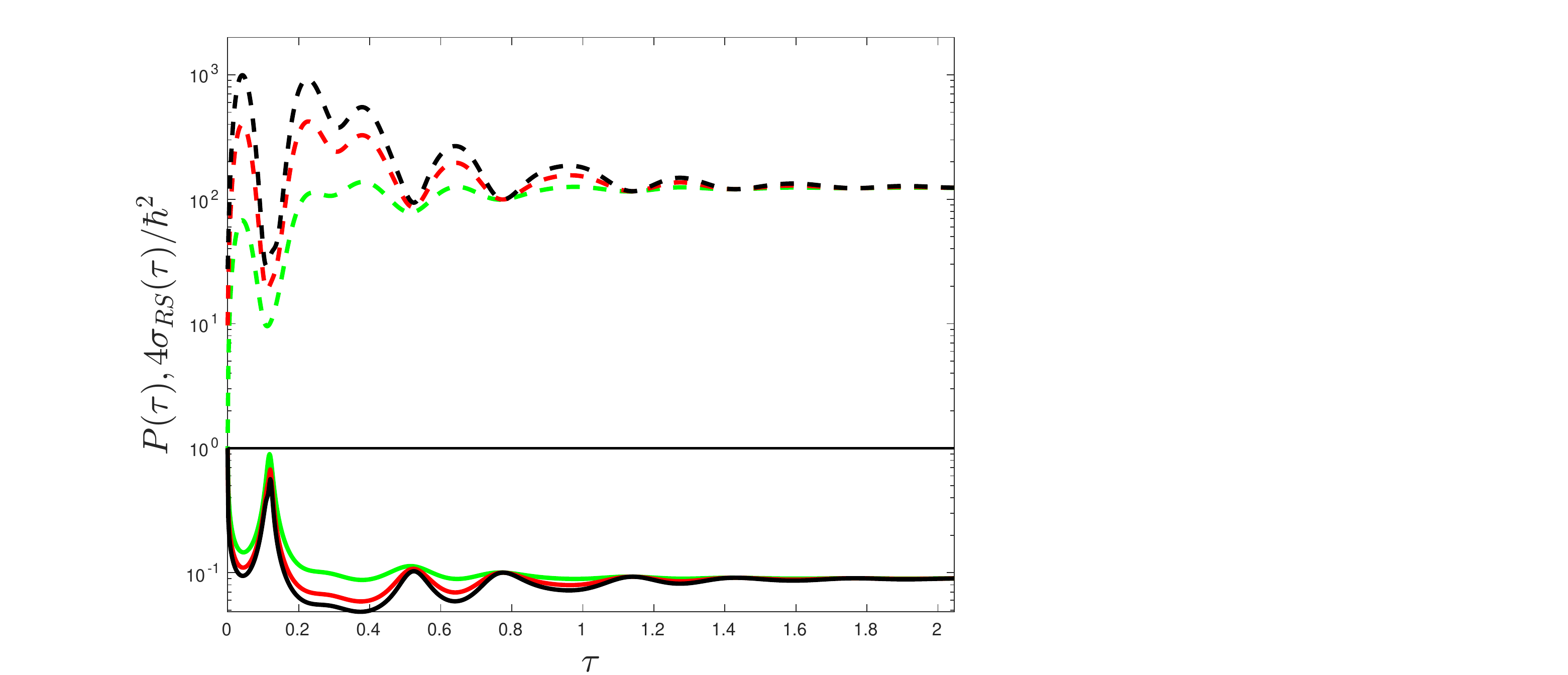}
			\caption{}
			
		\end{subfigure}
		\caption{Semilogarithmic plot of purity $P(\tau)$ (solid) and $4\sigma_{RS}(\tau)/\hbar^2$ (dashed) in Eq. \eqref{SRI} as a function of dimensionless time $\tau$. The parameters are  
			$\gamma'=0.15 $, $T'= 10\times  T'^{I}_{min}$ (see Eq. \eqref{Tmin1}), and $\beta'=0.6$ according to Eq. \eqref{transf}. Left panel: $D_{px}=0$ or case I in Table \ref{tab:table1}.
			Right panel: case II in Table \ref{tab:table1} with $\Omega'=1.25$ defined in Eq. \eqref{transf}. The solid black lines mark both the allowed upper bound for the purity and the right hand-side of 
			the Robertson-Schr\"odinger uncertainty relation. Both figures show that purity does not exceed one and the Robertson-Schr\"odinger uncertainty relation is not violated. 
			Here, we have a high enough temperature, which is always understood in the context of the inverse of the width of the initial wave packet $\beta'$. Furthermore, curves on both sides of the line at
			value one show some similar behavior in their slopes.} \label{fig2}
	\end{figure*}
	
	\section{Numerical results}
	\label{V}
	
	In this section we present the time evolution of the purity $P(t)$ by using the solution to the master equation in Sec. \ref{II} and Eq. \ref{Plan}. In parallel we are going to carry out a numerical
	investigation on the Robertson-Schr\"odinger uncertainty \eqref{SRI} with the position operator $\hat x$ and the momentum operator $\hat p$. This time we focus only on
	the two non-Lindblad master equations, cases I and II in Table \ref{tab:table1}, for a few interesting parameter constellations. The other two cases, where a Lindblad master equations is truncated
	to form of Eq. \eqref{MEQ}, are neglected due to our findings in Secs. \ref{III} and \ref{IV}.
	
	\begin{figure*}[ht!]
		\begin{subfigure}{.49\textwidth}
			\includegraphics[trim={8.5cm 20cm 0 20cm},clip,width=12cm]{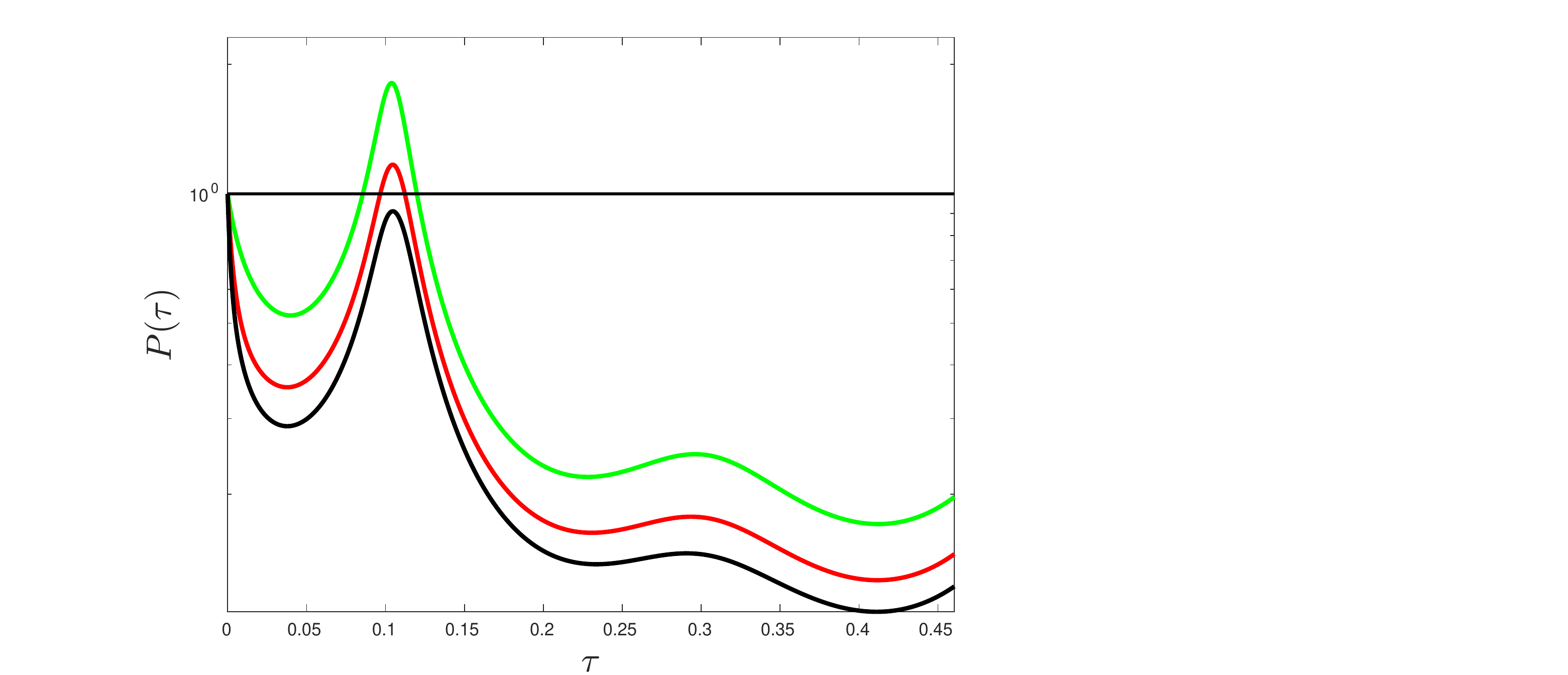}
			\caption{}
		\end{subfigure}
		\begin{subfigure}{.49\textwidth}
			\includegraphics[trim={8.5cm 20cm 0 20cm},clip,width=12cm]{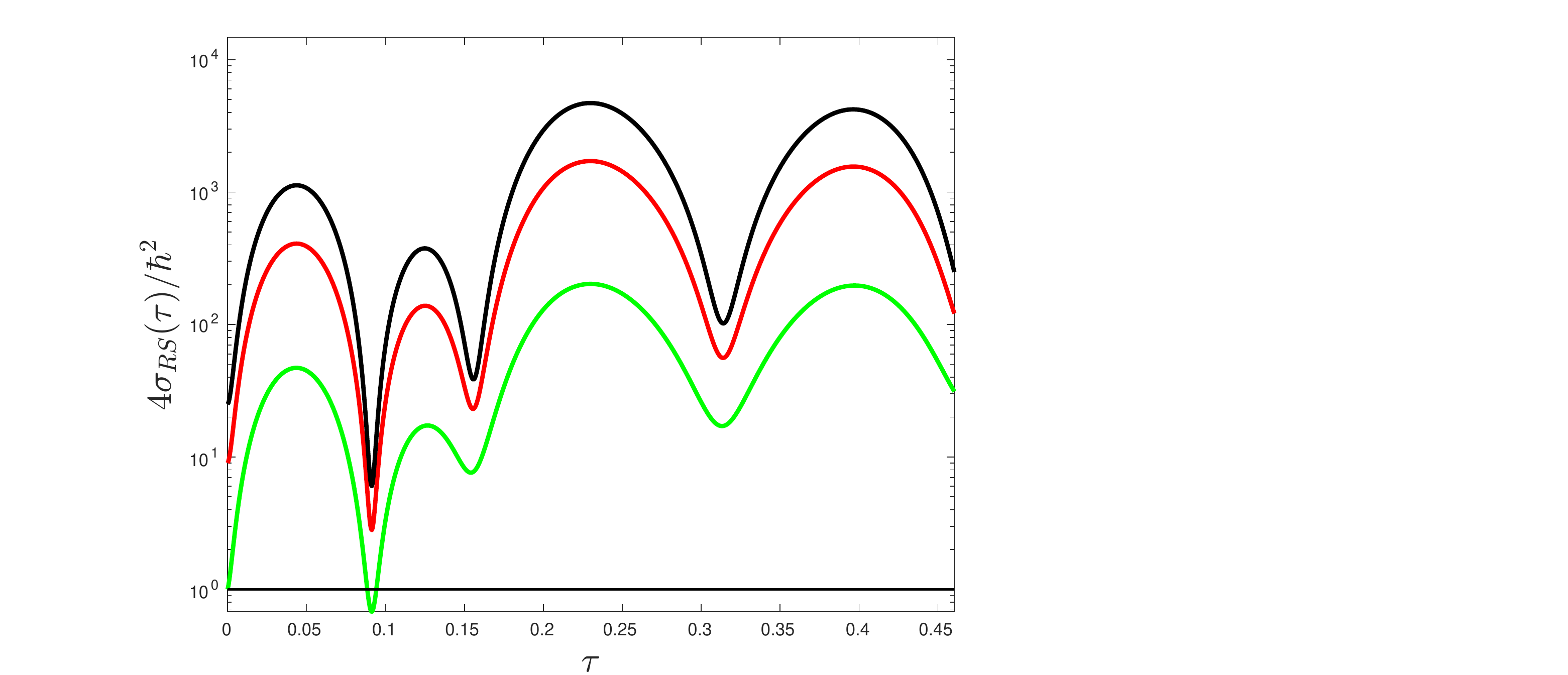}
			\caption{}
		\end{subfigure}
		\\
		\begin{subfigure}{.49\textwidth}
			\includegraphics[trim={8.5cm 20cm 0 20cm},clip,width=12cm]{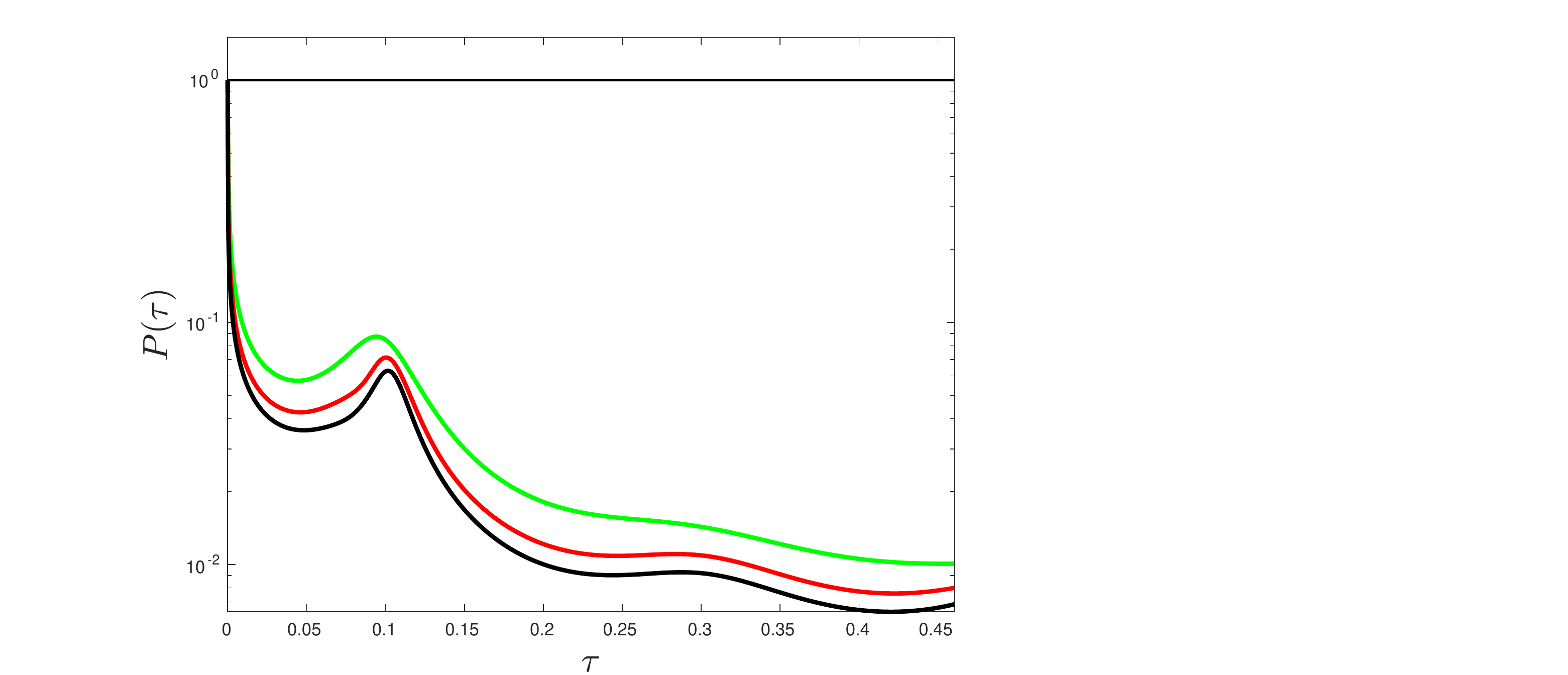}
			\caption{}
		\end{subfigure}
		\begin{subfigure}{.49\textwidth}
			\includegraphics[trim={8.5cm 20cm 0 20cm},clip,width=12cm]{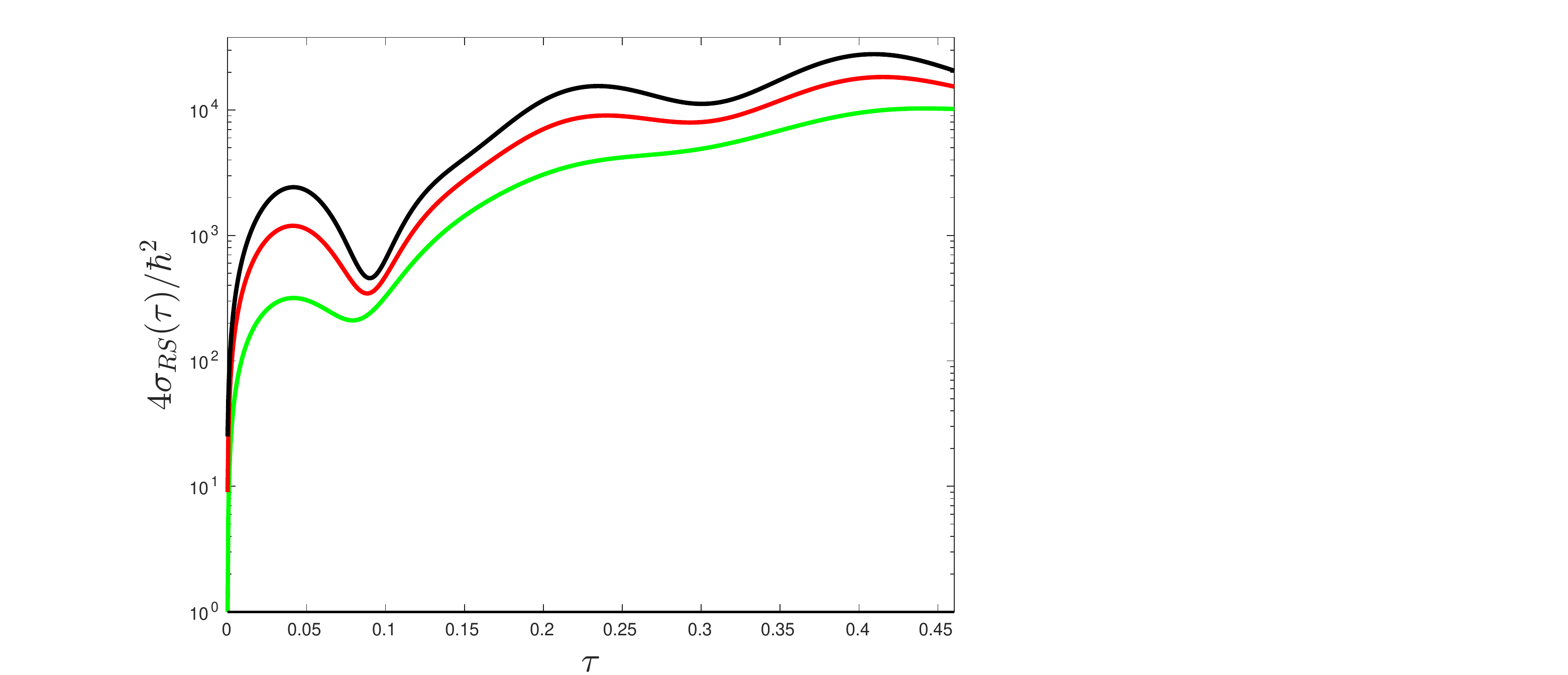}
			\caption{}
		\end{subfigure}
		\caption{Left panel: Semilogarithmic plot of purity $P(\tau)$ as a function of dimensionless time $\tau$. Right panel: Semilogarithmic plot of $4\sigma_{RS}(\tau)/\hbar^2$ in Eq. \eqref{SRI} as a function of $\tau$.
			The parameters are considered for case I in Table \ref{tab:table1}, i.e., $D_{px}=0$, $\gamma'=10^{-2}$, and $\beta'=0.6$ according to Eq. \eqref{transf}. Figures $a)$ and $b)$ are plotted for the same 
			temperature $T'=10 \times T'^{I}_{min}$ (see Eq. \eqref{Tmin1}). For figures $c)$ and $d)$ the temperature is set to $T'=10^3 \times T'^{I}_{min}$. In figures $a)$ and $b)$ violations with respect to the purity 
			and  $4\sigma_{RS}(\tau)/\hbar^2$ occur. Increasing the temperature these issues are resolved in $c)$ and $d)$.
			The solid black lines mark both the allowed upper bound for the purity and the right hand-side of the Robertson-Schr\"odinger uncertainty relation.}
		\label{fig3}
	\end{figure*}
	
	In Sec. \ref{IV} we have found that several initial pure states are subject to condition \eqref{cond} and therefore the purity will decrease from its initial value of one. Our purpose
	is to investigate whether the purity will exceed one at a later time with these particular pure initial states and when the Robertson-Schr\"odinger uncertainty is violated, i.e, $4 \sigma_{RS} < \hbar^2$. 
	We involve this extra task due to some previous investigations, where the positivity violation is investigated through the uncertainty principle, see for example \cite{Fleming}. We consider
	for our numerical investigations the following pure states ($ n=0,1,2$)  and in
the representation used in Sec. \ref{II} they read
	\begin{eqnarray}
	&&\rho(R,r,0)= 
	\frac {H_n \left(\beta R+\frac{\beta r}{2} \right) H_n\left(\beta R-\frac{\beta r}{2}\right)}{{2}^{n}n!} \times \nonumber \\ &&e^{-\frac{1}{2}\,{\beta}^{2}
		\left( 2\,{R}^{2}+\frac{1}{2}\,{r}^{2} \right) }\sqrt {{\frac {{
					\beta}^{2}}{\pi }}},\label{hos}
	\end{eqnarray}
	where we have used the center of mass and relative coordinates $R=(x+y)/2$, $r=x-y$. $H_n(x)$ are the Hermite polynomials. The parameter $\beta$ is proportional to 
	the inverse width $x_0=\sqrt{\hbar / (m \omega)}$ of the quantum harmonic oscillator's ground state. Furthermore,
	we introduce the following dimensionless parameters:
	\begin{eqnarray}
	&&D'_{pp}=\frac{D_{pp} x^2_0}{\omega}, \quad D'_{px}=D_{px}m x^2_0,\quad \gamma'=\frac{\gamma}{\omega}, \nonumber \\
	&&\Omega'=\frac{\Omega}{\omega},\quad \beta'=\beta x_0,\quad   \label{transf}  
	\tau= \omega t, \quad T'=\frac{k_B T}{\hbar \omega}.
	\end{eqnarray}
	We note, that in case of $\beta'=\beta x_0=1$ these states are the eigenstates of 
	the quantum harmonic oscillator. It is apparent from Fig. \ref{fig1} that pure states subject
	to condition \eqref{cond} prescribe a short initial time evolution, where purity does not exceed one and the Robertson-Schr\"odinger uncertainty relation is also not violated. 
	
	We argued in Sec. \ref{I} about the relation between the purity and the inequality of Robertson-Schr\"odinger uncertainty principle. In Fig. \ref{fig2} we see that for high enough temperatures there is a striking 
	analogous 
	behavior in both cases, namely the slopes of the plotted curves are almost identical. Motivated by this fact we have carried out a brief calculation in Appendix \ref{AppII}, which shows that whenever
	the derivative of the purity with respect to time is zero it corresponds in most of the cases to situations where the time derivative of $4 \sigma_{RS}$ is also zero. However, there are some cases when this is not 
	true and they belong to situations where violations with respect to purity or the Robertson-Schr\"odinger uncertainty relation occur.
	
	\begin{figure*}[ht!]
		\begin{subfigure}{.49\textwidth}
			\includegraphics[trim={8.5cm 20cm 0 20cm},clip,width=12cm]{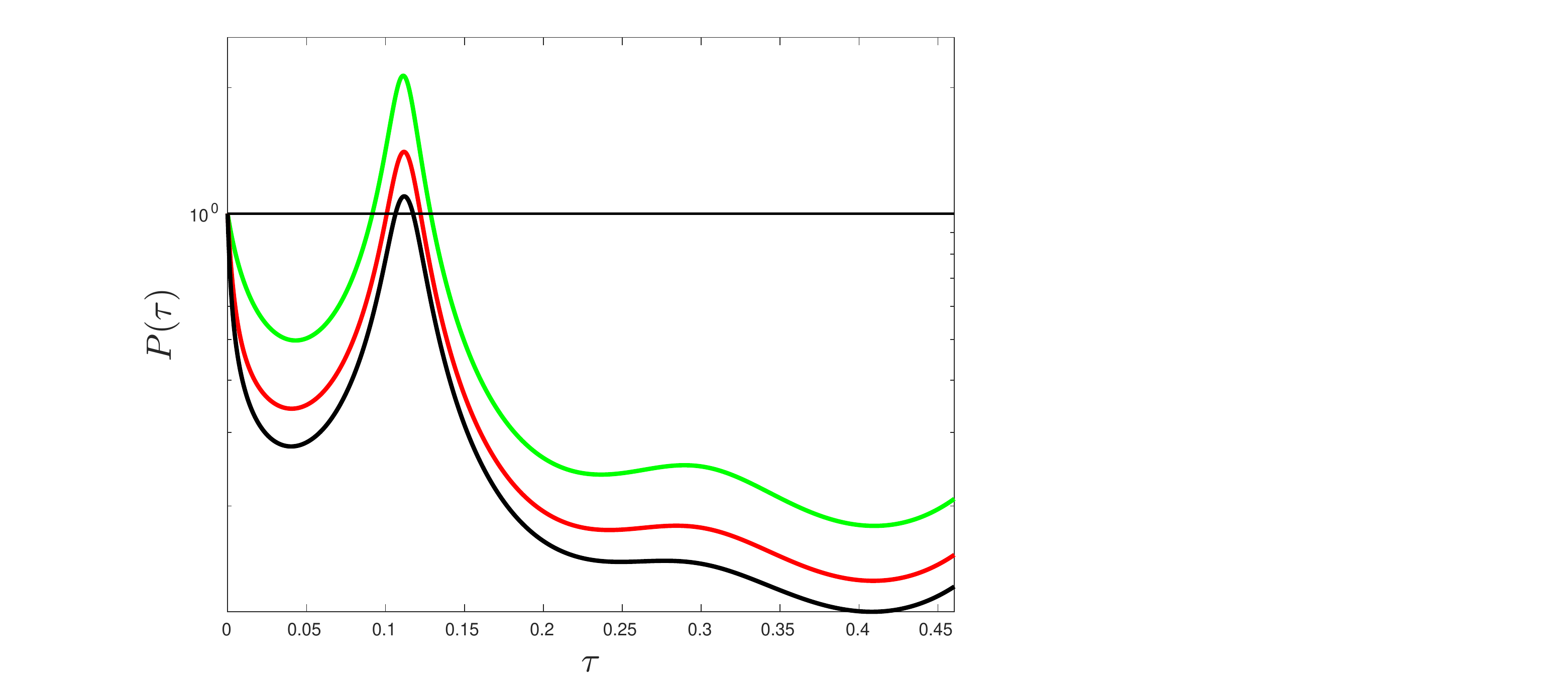}
			\caption{}
		\end{subfigure}
		\begin{subfigure}{.49\textwidth}
			\includegraphics[trim={8.5cm 20cm 0 20cm},clip,width=12cm]{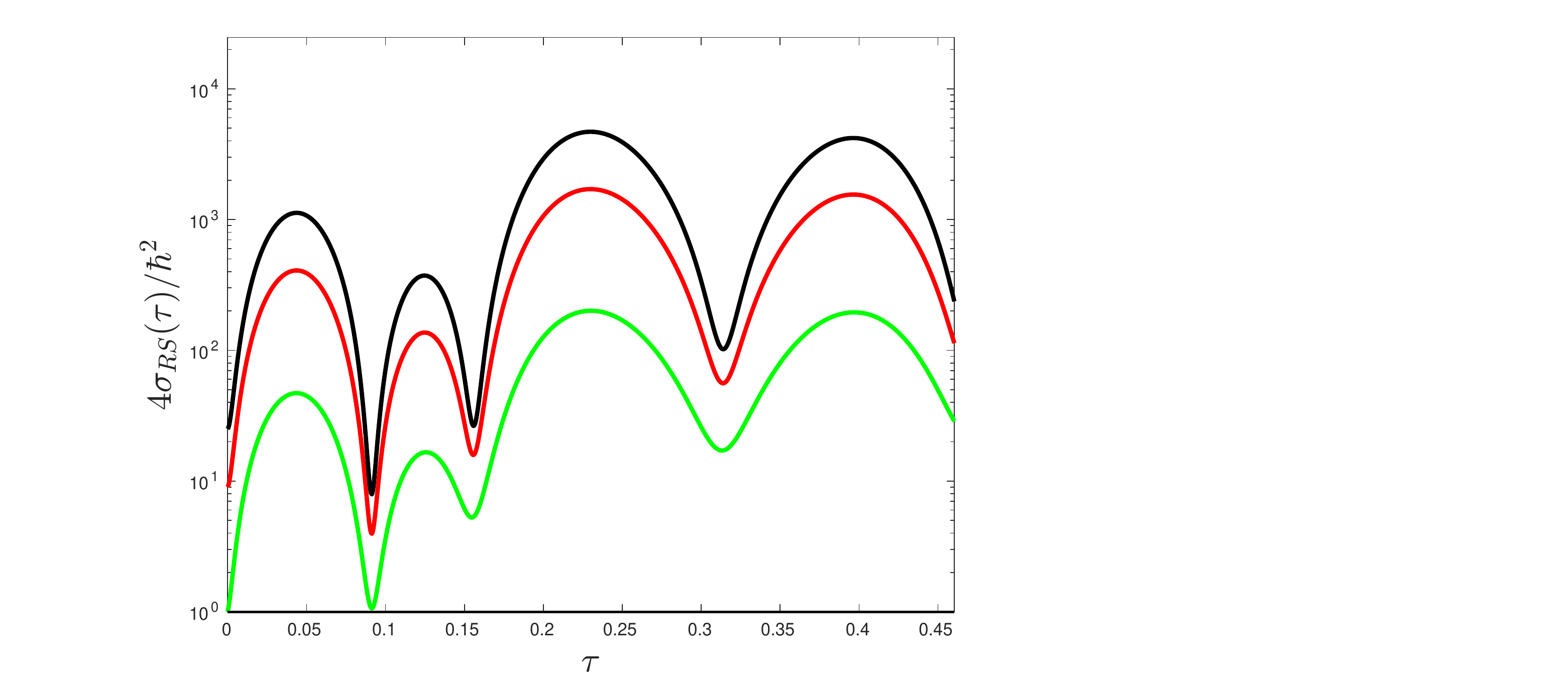}
			\caption{}
		\end{subfigure}
		\\
		\begin{subfigure}{.49\textwidth}
			\includegraphics[trim={8.5cm 20cm 0 20cm},clip,width=12cm]{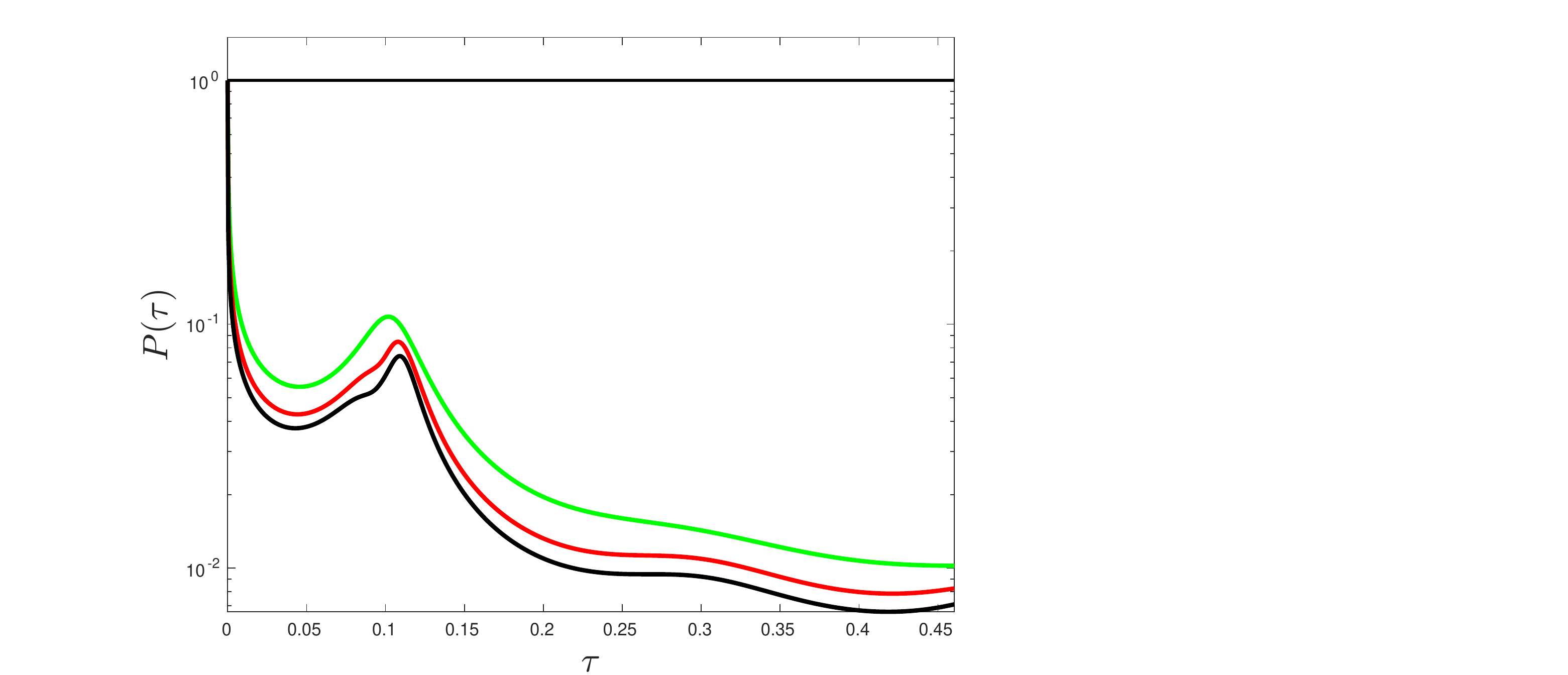}
			\caption{}
		\end{subfigure}
		\begin{subfigure}{.49\textwidth}
			\includegraphics[trim={8.5cm 20cm 0 20cm},clip,width=12cm]{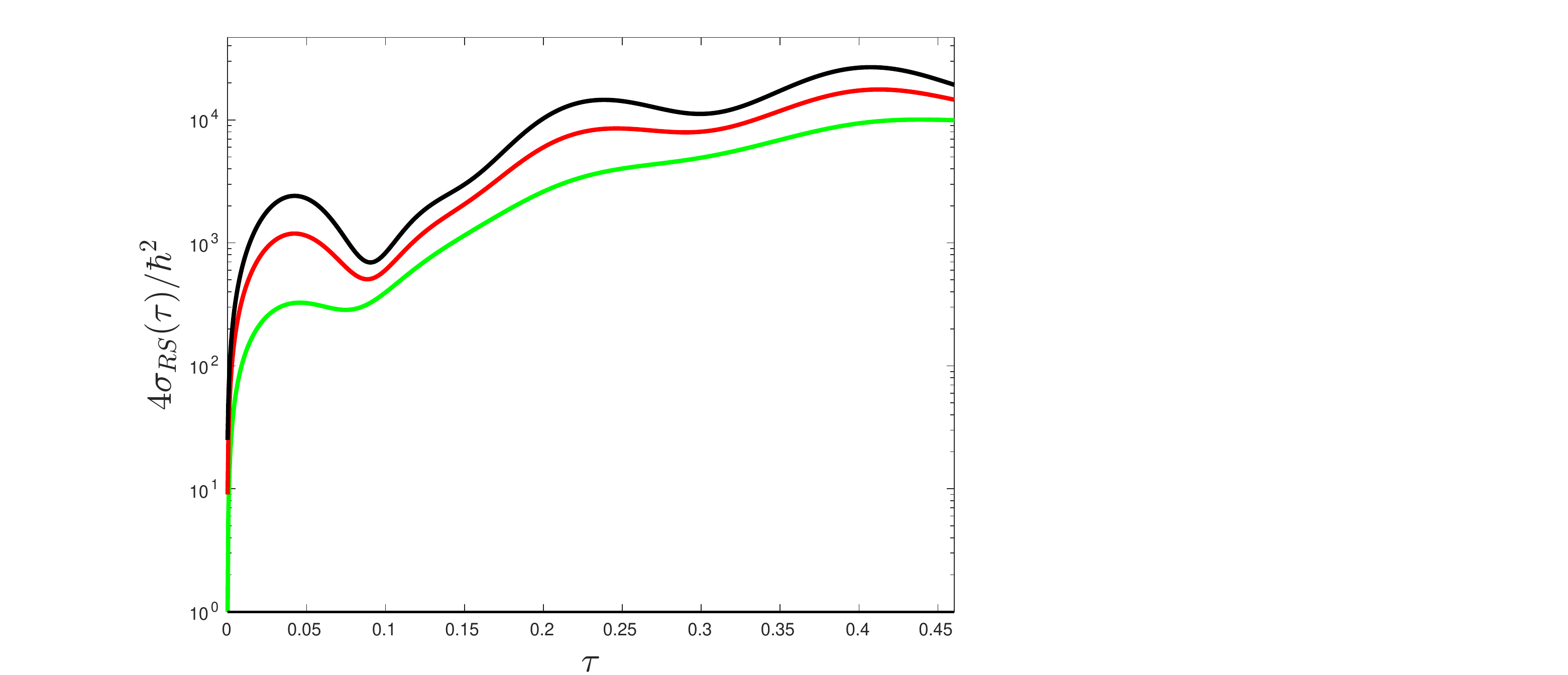}
			\caption{}
		\end{subfigure}
		\caption{Left panel: Semilogarithmic plot of purity $P(\tau)$ as a function of dimensionless time $\tau$. Right panel: Semilogarithmic plot of $4\sigma_{RS}(\tau)/\hbar^2$ in Eq. \eqref{SRI} as a function of $\tau$.
			The parameters are considered for case II of Table \ref{tab:table1}: $\Omega'=2$, $\gamma'=10^{-2} $ and $\beta'=0.6$ according to Eq. \eqref{transf}. The temperatures are set for figures $a)$, $b)$, 
			$c)$ and $d)$ to the same value as in Fig. \ref{fig3}. Small width of the initial wave packets result in violations with respect to the purity and  $4\sigma_{RS}(\tau)/\hbar^2$.
			Increasing the temperature resolves again these issues. The solid black lines mark both the allowed upper bound for the purity and the right hand-side of the Robertson-Schr\"odinger uncertainty relation.}
		\label{fig4}
	\end{figure*}
	
	According to condition \eqref{cond}, we have considered not too high temperatures due to the small width of the initial wave packet, but they are still way above $T'^I_{min}$ which has been determined in
	Sec. \ref{III} in the context of the steady state. In numbers, the first temperature choice is $T_1=10 \times T'^I_{min}$ or $k_B T_1=20 \hbar \omega$ and the second one
	$T_2=10^3 \times T'^I_{min}$ or $k_B T_2=2000 \hbar \omega$, thus both temperatures fulfill the high temperature approximation $k_B T_{1,2} \gg \hbar \omega$ employed in the derivation
	of the master equation. Figs. \ref{fig3} and \ref{fig4} show that initial states which are subject to the condition \eqref{cond} may lead to the violation of the positivity of the 
	density operator. As the temperature is increased these issues are resolved, a demonstration of our analytical results on the derivative of the purity with respect to the temperature in Sec. \ref{III}. 
	We can also see violations which do not occur in the same time in purity $P(\tau)$ and $4\sigma_{RS}(\tau)/\hbar^2$, the left hand-side of the Robertson-Schr\"odinger uncertainty relation in \eqref{SRI}, 
	thus Appendix \ref{AppII} cannot cover these cases analytically. In other words, the sets $\mathcal{D}^P_1(\mathcal{H})$ and $\mathcal{D}^I_1(\mathcal{H})$, introduced in Sec. \ref{I}, do not overlap at all times. 
	Furthermore, the numerical results show that somehow the purity is more sensitive than the Robertson-Schr\"odinger uncertainty relation to the positivity violation of the density operator. In summary, we have 
	been able to demonstrate numerically that initial states fulfilling condition \eqref{cond} still might lead to the positivity violation of the density operator in the later time of evolution.

	\section{Discussion and conclusions }
	
	We have solved analytically the Caldeira-Leggett master equation of a quantum harmonic oscillator with the method of characteristics curves. Our choice on the method of characteristics is motivated
	by our focus on positivity violations of the density operator. The out most goal would be to find all initial states for which there are times during the 
	time evolution such that there exists at least one negative eigenvalue of the master equation's solution. In general this is not a simple mathematical task, definitely not for the infinite dimensional
	Hilbert space of the quantum harmonic oscillator. Therefore, we have taken the purity of the states, but we have to remind the reader about the fact that the purity being between 
	the values one and zero is a necessary but not sufficient condition that there is 
	no positivity violation during the time evolution. 
	
	As a first task, we have investigated the steady state solution and we have considered four cases for the values of the diffusion coefficients $D_{pp}$ and $D_{px}$. These cases consist of: the pioneering work of 
	Caldeira and Leggett \cite{CL}; an extended derivation of their result in \cite{book1}; another extension \cite{DIOSI1993517}, where the derivation drops the high temperature limit and focuses on medium 
	temperatures; and finally the master equation obtained via a phenomenological phase space quantization of an under-damped harmonic oscillator \cite{dekker}. The last two cases deal with Lindblad master equations, 
	where positivity violations cannot occur due to the form of mathematical map on the density operators. However, we have truncated these master equations in order to obtain a Caldeira-Leggett master equation. 
	In the first three cases the investigation on the steady states led to some conditions upon the temperature which are in accordance with the approximations used for the derivation of the master equations. In the 
	last case, it turned out that the master equation of Dekker \cite{dekker} derived for zero temperature cannot be truncated at will. We have also shown the derivative of the purity with respect to the 
	temperature is always smaller than zero. Hence, situations where the purity exceeds one can be resolved by increasing the temperature. Thus, we have concluded: 
	any kind of positivity violations of the density operator have to be searched for short and intermediate evolution times.
	
	Therefore, in the next task we have focused on cases where the purity is one at a certain time. We have found the requirements on the width of the wave packet such that the time derivative of the purity 
	is not positive, i.e, the purity decreases from the value one. In fact, the most natural way to apply this result is for initial pure states. In the case of the parameters given in the work of 
	Caldeira and Leggett the initial
	width of the wave packet has to be larger than five times the thermal wavelength, which does not apply to many pure states. This statement suggests that one may find short time evolutions of the 
	Caldeira-Leggett master equation for certain initial states to be mathematically inconsistent. Due to the steady state positivity, these initial problems disappear and in fact they can be ignored. This argument 
	is more or less
	known in the community, however based on our full analytical knowledge on the evolving state and the exact details on the width of the wave packet, we have considered for initial conditions a set of pure 
	states including also the first three eigenstates of the 
	quantum harmonic oscillator. While they fulfill the condition on their wave packet width, at later times of the evolution the positivity violation of the density operator can be found. To show
	it, we have carried out numerical simulations and indeed we have been able to show that the purity exceeds one for later times. In the numerical investigations we have also compared the behavior of the purity and the Robertson-Schr\"odinger uncertainty relation, where the latter one
	is mostly preferred as a test for the positivity violation \cite{Fleming}. We have found a remarkable agreement between the slopes of their curves. However, when the positivity violation occurs, the purity
	seems to be a more sensitive indicator for the existence of any negative eigenvalue. As this observation is made with the help of numerical investigations the analytical proof is still missing and 
	is subject of ongoing investigations. 
	
	Finally let us make some comments on our results. Both the analytical and numerical results clearly indicate inconsistencies in the Caldeira-Leggett master equation. 
	We have reobtained some known facts, but also some new ones like the situations with initial pure states in Eq. \ref{hos}. One fact is clear, if this Markovian master equation with high enough temperature is applied for 
	long times of evolution, then effective and desirable descriptions of certain physical systems can be obtained. For example, in optomechanics, where quantum effects at low temperatures are important the use of 
	the Caldeira-Leggett master equation may lead to inconsistencies, a fact which has initiated extended studies on quantum dissipative models for harmonic oscillators \cite{Ref_10,Ref_11}. However, 
	the requirement for mathematical consistency is still looming over the master equation and there might be other surprising situations than those presented in this work, where the positivity of the density 
	operator is violated not only for short time evolutions.

	\section*{Acknowledgements}
	The authors have profited from helpful discussions with M. A. Csirik, Z. Kaufmann, G. Helesfai, L. Di\'osi, A. Csord\'as, T. Geszti, Z. Zimbor\'as, and G. Csizmadia. 
	This research is supported by the National Research Development and Innovation Office of Hungary within the Quantum Technology National Excellence Program (Project No. 2017-1.2.1-NKP-2017-00001) and
	the European Union's Horizon 2020 research and innovation programme under Grant Agreement No. 732894 (FET Proactive HOT).
	
	\section*{Author contribution statement}
	J. Z. B. and G. H. conceived of the presented idea. G. H. and J. Z. B. developed the theory. G. H. and L. L. performed the computations.
	All authors verified the analytical methods, discussed the results and contributed to the final manuscript.

	\appendix
	\section{Detailed expressions of the introduced notations}
	\label{AppI}
	
	In this Appendix, we present the full expressions of several notations introduced in the main text. First, the notations supporting Eq. \eqref{solution} read
	\begin{eqnarray}
	&&A:=2\,{X}^{3/2} \left( {\omega}^{2}{r}^{2}+{K}^{2} \right) {e^{2\,t
			\left( 2\, \sqrt{X}+\gamma \right) }}+ \Big(  \left( -\gamma\,{
		\omega}^{2}{r}^{2}
	\right. \nonumber\\
	&& \left.
	+  2\,K{\omega}^{2}r-{K}^{2}\gamma \right)  \sqrt{X}+X
	\left( -{\omega}^{2}{r}^{2}  +{K}^{2} \right)  \Big) \gamma\,{e
		^{2\,t \sqrt{X}}} \nonumber \\
	&&+2\,{e^{4\,t \sqrt{X}}}{\omega}^{2} \sqrt{X}
	\left( {\omega}^{2}{r}^{2}-2\,K\gamma\,r+{K}^{2} \right)-{e^{6
			\,t \sqrt{X}}} \times \nonumber \\
	&&\Big(  \left( \gamma\,{\omega}^{2}{r}^{2}-2\,K{\omega}
	^{2}r+{K}^{2}\gamma \right) \sqrt{X}+X \left( -{\omega}^{2}{r}^{2}+{K
	}^{2} \right)  \Big) \gamma, \nonumber
	\end{eqnarray}
	and
	\begin{eqnarray}
	&&B:=-8\,{X}^{3/2}\gamma\,{K}^{2}{e^{2\,t \left( 2\, \sqrt{X}+\gamma
			\right) }}+ \Bigg(  \Big(  \left( 4\,{\gamma}^{2}-2\,{\omega}^{2}
	\right) {K}^{2}- \nonumber \\
	&& 4\,{\omega}^{2}r\gamma\,K+2\,{\omega}^{4}{r}^{2}
	\Big)  \sqrt{X}+  
	X \left( 4\,K{\omega}^{2}r-4\,{K}^{2}\gamma
	\right)  \Bigg) \gamma\,{e^{2\,t \sqrt{X}}} \nonumber \\
	&&-4\,{e^{4\,t
			\sqrt{X}}}{\omega}^{2} \sqrt{X} \left( {\omega}^{2}{r}^{2}-2\,K\gamma
	\,r+{K}^{2} \right) \gamma-{e^{6\,t \sqrt{X}}}  \times \nonumber \\ 
	&& 
	\Bigg(  \left[
	\Big( -4\,{\gamma}^{2} +2\,{\omega}^{2} \Big) {K}^{2}+4\,{\omega}^{
		2}r\gamma\,K-2\,{\omega}^{4}{r}^{2} \right]  \sqrt{X} \nonumber \\
	&&+X \left( 4\,K{
		\omega}^{2}r-4\,{K}^{2}\gamma \right)  \Bigg) \gamma. \nonumber
	\end{eqnarray}
	
	Secondly, the specific functions of Sec. \ref{III} read
	\begin{eqnarray}
	&&G(K,r,t)=-\Re \left[ \frac{T}{{ X^{\frac{3}{2}}\Omega\,{\omega}^{2}}} \left[ -\frac{1}{2}\,\gamma\, \left(  \left( -2\,{K}^{2}{
		\gamma}^{2} \right. \right. \right. \right. \nonumber  \\ 
	&&\left. \left. \left. \left. + \left(  \left( \Omega\,{r}^{2}+2\,Kr \right) {\omega}^{2}
	+{K}^{2}\Omega \right) \gamma+ \left( -{\omega}^{2}{r}^{2}+K \left( -2
	\,\Omega\,r   \right. \right.\right. \right. \right. \right. \nonumber  \\
	&& \left. \left. \left. \left. \left. \left.+K \right)  \right) {\omega}^{2} \right) \sqrt {X}+ \left( -2\,{K}^{2}\gamma+ \left( -\Omega\,{r}^{2}+2\,
	Kr \right) {\omega}^{2}  \right. \right. \right. \right. \nonumber  \\
	&&\left. \left. \left. \left.
	+ {K}^{2}\Omega \right)  X  \right) e^{4\,t\sqrt{X}}+{\omega}^{2}\sqrt {X} \left( {\omega}^{2}{r}^{2} \right. \right. \right.  \nonumber  \\
	&& \left. \left. \left. -2\,\gamma\,Kr+{K}^{2} \right)  \left( 
	\Omega-\gamma \right) e^{2\,t\sqrt {X}
	}+\frac{1}{2}\, \left(  \left( 2\,{K}^{2}{\gamma}^{2}+ \right. \right. \right. \right. \nonumber \\
	&& \left. \left. \left. \left. \left(  \left( -\Omega
	\,{r}^{2}-2\,Kr \right) {\omega}^{2}-{K}^{2}\Omega \right) \gamma-
	\left( -{\omega}^{2}{r}^{2} \right. \right. \right. \right. \right. \nonumber \\
	&& \left. \left. \left. \left. \left. +K \left( -2\,\Omega\,r+K \right) 
	\right) {\omega}^{2} \right) \sqrt {X}  +
	\left( -2\,{K}^{2}\gamma+ \left(-\Omega\,{r}^{2} \right. \right. \right. \right. \right. \nonumber \\ 
	&& \left. \left. \left. \left. \left. +2\,Kr \right) {
		\omega}^{2}+{K}^{2}\Omega \right)  X \right) \gamma \right] {e^{-2\,t
			X }} \right], \nonumber
	\end{eqnarray}
	and
	\begin{eqnarray}
	&&F(K,r,t)=\Re \left[ \frac{1}{2\left({{X}^{\frac{3
				}{2}}{T}^{2}{\omega}^{2}\pi }\right)}\, \left[ \gamma\, \left(  \left( \frac{1}{3}\,{K}^{2}\Omega\,{
		\gamma}^{2} \right.\right. \right. \right. \nonumber  \\
	&& \left. \left. \left. \left.+ \left(  \left( \pi \,{T}^{2}{r}^{2}-\frac{1}{3}\,Kr\Omega
	\right) {\omega}^{2}+\pi \,{T}^{2}{K}^{2} \right) \gamma\pi \,{T}^{2}{K}^{2} \right) X \right) \times   \right. \right. \nonumber  \\
	&& \left. \left.
	\times{e^{2\,t\sqrt {X}}}+
	\left(  \left( \frac{1}{3}\,{K}^{2}\Omega\,{\gamma}^{2}+ \left(  \left( \pi 
	\,{T}^{2}{r}^{2}-\frac{1}{3}\,Kr\Omega \right) {\omega}^{2}+ \right. \right. \right. \right. \right. \nonumber  \\ 
	&&\left.\left. \left. \left. \left. 
	K \left( 12\,\pi \,{T}^{2}r+K\Omega \right)  \right)  \right) \sqrt 
	{X}+ \left( \frac{1}{3}\,{K}^{2}\Omega\,\gamma+ \left( -\pi \,{T}^{2}{r}^{2} 
	\right. \right. \right. \right. \right.  \nonumber  \\
	&& \left. \left. \left. \left. \left.  -\frac{1}{3}
	\,Kr\Omega \right) {\omega}^{2}+\pi \,{T}^{2}{K}^{2} \right) X
	\right) \gamma\,{e^{6\,t\sqrt {X}}} -2\,\sqrt {X}{\omega}^{2} \times  \right. \right. \nonumber \\
	&&\left. \left.  \left( {\omega}^{2}{r}^{2}  -2\,K\gamma\,r+  {K}^{2} \right) {e^{4
			\,t\sqrt {X}}} \left( \pi \,{T}^{2}+  
	\frac{1}{6}\,\gamma\,\Omega \right) 
	\right] \times \right. \nonumber \\
	&& \left. {e^{-2\,t \left( 2\,\sqrt {X}+\gamma \right) }}{{X}^{\frac{3
			}{2}}{T}^{2}{\omega}^{2}\pi } \right], \nonumber
	\end{eqnarray}
	where $\Re$ stands for the real part of a complex number.
	
	\section{Existence proof about common extrema in the time evolution of $\sigma_{RS}(t)$ and $P(t)$ }
	\label{AppII}
	First we calculate the  expression for the derivative of the Robertson-Schr\"odinger formula with respect to 
	time in a Fourier transformed representation. The suitable  moments of position and momentum operators in this representation are
	\begin{eqnarray}
	\left\langle \hat{x}\right\rangle =\frac{i}{\sqrt {2\pi }}\int_{-\infty }^{\infty }\!\int_{-\infty }^{\infty }\!{
		{\rm e}^{iKR}} {\frac {\partial   \rho \left( K,0,t \right)   }{\partial  K  }}\,{\rm d}K\,{\rm d}R, \nonumber
	\end{eqnarray}
	
	\begin{eqnarray}
	\left\langle \hat{x}^2\right\rangle =-\frac{1}{\sqrt {2\pi }}\int_{-\infty }^{\infty }\!\int_{-\infty }^{\infty }\!{
		{\rm e}^{iKR}} {\frac {\partial^2   \rho \left( K,0,t \right)   }{\partial  K^2 }}\,{\rm d}K\,{\rm d}R, \nonumber
	\end{eqnarray}
	
	\begin{eqnarray}
	&& \left\langle \hat{p} \right\rangle =\left.-\frac{i}{\sqrt {2\pi }}\int_{-\infty }^{\infty }\!\int_{-\infty }^{\infty }\!{
		{\rm e}^{iKR}} \right. \nonumber \\
	&& \left. \times \left[\frac{i}{2}K+\frac{\partial }{\partial r} \right]\rho(K,r,t)\right|_{r=0}  \,{\rm d}K\,{\rm d}R, \nonumber
	\end{eqnarray}
	
	\begin{eqnarray}
	&& \left\langle \hat{p}^2 \right\rangle =\left.\frac{1}{\sqrt {2\pi }}\int_{-\infty }^{\infty }\!\int_{-\infty }^{\infty }\!{
		{\rm e}^{iKR}} \right. \nonumber \\
	&& \left. \times \left\lbrace \left[\frac{1}{4}K^2+ i K\frac{\partial }{\partial r}-\frac{\partial^2 }{\partial r^2} \right]\rho(K,r,t)\right\rbrace \right|_{r=0}  \,{\rm d}K\,{\rm d}R, \nonumber
	\end{eqnarray}
	
	\begin{eqnarray}
	&& \frac{1}{2}\left\langle \hat{x}\hat{p}+\hat{p}\hat{x}  \right\rangle =\left.\frac{1}{\sqrt {2\pi }}\int_{-\infty }^{\infty }\!\int_{-\infty }^{\infty }\!{
		{\rm e}^{iKR}} \right. \nonumber \\
	&& \left. \times \left\lbrace  \left[-\frac{1}{2}K\frac{\partial }{\partial K}+ \frac{Kr}{4}+\frac{\partial^2}{\partial r \partial K}+\frac{r}{2} \frac{\partial}{\partial r} \right]\rho(K,r,t)\right
	\rbrace \right|_{r=0}  \nonumber \\
	&& \times \,{\rm d}K\,{\rm d}R, \nonumber
	\end{eqnarray} 
	and we finally  get
	\begin{eqnarray}
	&&\frac{d \sigma_{RS}(t)}{dt}=\left[ \frac{d}{dt}\left\langle \hat{x}^2\right\rangle-2\left\langle \hat{x}
	\right\rangle\frac{d}{dt}\left\langle \hat{x}\right\rangle \right]\cdot\left[\left\langle \hat{p}^2\right\rangle-\left\langle \hat{p}\right\rangle^2 \right] \nonumber \\
	&&+ \left[ \frac{d}{dt}\left\langle \hat{p}^2\right\rangle-2\left\langle \hat{p}
	\right\rangle\frac{d}{dt}\left\langle \hat{p}\right\rangle \right]\cdot\left[\left\langle \hat{x}^2\right\rangle-\left\langle \hat{x}\right\rangle^2 \right]  \nonumber \\
	&& - 2\left[\frac{1}{2}\left\langle \hat{x} \hat{p}+\hat{p} \hat{x}\right\rangle-\left\langle \hat{x}\right\rangle\left\langle \hat{p}\right\rangle \right] \nonumber \\
	&&\times \left(\frac{1}{2}\frac{d}{dt}\left\langle \hat{x} \hat{p}+\hat{p} \hat{x}\right\rangle-\left\langle 
	\hat{p} \right\rangle \frac{d}{dt}\left\langle \hat{x} \right\rangle-\left\langle \hat{x} \right\rangle \frac{d}{dt}\left\langle \hat{p} \right\rangle \right).
	\nonumber
	\end{eqnarray}
	The purity in the Fourier transformed representation reads
	\begin{eqnarray}
	&&\frac{d}{d t}P(t)=\int_{-\infty}^{\infty}{\int_{-\infty}^{\infty}{\rho(K,r,t)\frac{\partial}{\partial t}\rho^{\ast}(K,r,t) dK dr}} \nonumber \\
	&&+\int_{-\infty}^{\infty}{\int_{-\infty}^{\infty}{\rho^{\ast}(K,r,t)\frac{\partial}{\partial t}\rho(K,r,t) dK dr}}.
	\nonumber 
	\end{eqnarray}
	Let us suppose that there are times $t^*$, when
	\begin{equation}
	\left. \frac{\partial \rho(K,r,t)}{\partial t} \right|_{t=t^*}=0 \rightarrow \left. \frac{\partial \rho^{\ast}(K,r,t)}{\partial t}\right|_{t=t^*}=0, \nonumber
	\end{equation}
	because the  real and imaginary parts of derivative of the density function with respect to time must disappear in the same time. 
	If the  necessary mathematical conditions are satisfied, the derivation and the integration can be exchanged,
	then the following mathematical identity is fulfilled:
	
	\begin{eqnarray}
	&&\frac{d}{d t} \int_{-\infty}^{\infty}{\int_{-\infty}^{\infty}\mathcal{O}(K,r)\rho(K,r,t) dK dr}= \nonumber \\
	&&\int_{-\infty}^{\infty}\int_{-\infty}^{\infty}\mathcal{O}(K,r)\frac{\partial}{\partial t}\rho(K,r,t) dK dr, \nonumber
	\end{eqnarray}
	where $\mathcal{O}(K,r)$ is the corresponding representation of an operator $\hat{\mathcal{O}}$.
	
	Hence, there are times $t^*$ when the time derivative of both functions $\sigma_{RS}(t)$ and $P(t)$ disappear, since in this case the constant zero function is integrated in both cases.


\begin{thebibliography}{99}
		
		\bibitem{Neumann}
		%
		J. von Neumann, {\it Mathematische Grundlagen der Quantenmechanik} (Springer-Verlag, Berlin, 1932).
		%
		\bibitem{GKAS} V. Gorini, A. Kossakowski and E. C. G. Sudarshan, J. Math. Phys. {\bf 17}, 821 (1976).
		%
		\bibitem{lindblad1976} G. Lindblad, Comm. Math. Phys. {\bf 48}, 119 (1976).
		%
		\bibitem{Ref_3}
		A. O. Caldeira and A. J. Leggett,
		Ann. Phys. (N.Y.) {\bf 149}, 374 (1983).
		%
		\bibitem{CL} A. O. Caldeira and A. J. Leggett, Physica {\bf 121A}, 587 (1983).
		%
		%
		\bibitem{Spohn} H. Spohn, Rev. Mod. Phys. {\bf 52}, 569 (1980).
		%
		\bibitem{Talkner}
		P. Talkner,
		Ann. Phys. {\bf 167}, 390 (1986).
		%
		\bibitem{Ref_4} D. Kohen, C. C. Marston and D. J. Tannor, J. Chem. Phys. {\bf 107}, 5236 (1997).
		%
		\bibitem{Kraus1} K. Kraus, Ann. Phys. {\bf 64}, 311 (1971).
		%
		\bibitem{book1} H.-P. Breuer and F. Petruccione, {\it The theory of open quantum systems} (Oxford University Press, Oxford, 2002).
		%
		\bibitem{Ref_1}
		G. Lindblad,
		Rep. Math. Phys. {\bf 10}, 393 (1976).
		%
		\bibitem{DIOSI1993517} L. Di\'{o}si, Physica A {\bf 199}, 517 (1993).
		%
		\bibitem{dekker} H. Dekker, Phys. Rep. {\bf 80}, 1 (1981).
		%
		\bibitem{Ref_6}
		A. Sandulescu and H. Scutaru,
		Ann. Phys. (N.Y.) {\bf 173}, 277 (1987).
		%
		\bibitem{Ref_7}
		L. Di\'{o}si,
		Europhys. Lett. {\bf 22}, 1 (1993).
		%
		\bibitem{Ref_8}
		J. G. Peixoto de Faria and M. C. Nemes,
		J. Phys. A: Math. Gen. {\bf 31}, 7095  (1998).
		%
		\bibitem{Ref_9}
		F. Petruccione and	B. Vacchini,
		Phys. Rev. E {\bf 71}, 046134 (2005).
		%
		\bibitem{Ref_10}
		A. Barchielli and	B. Vacchini,
		New J. Phys. {\bf 17}, 083004 (2015).
		%
		\bibitem{Gnutzmann1996} S. Gnutzmann and F. Haake, Z. Phys. B {\bf 101}, 263 (1996).
		%
		\bibitem{RV} S. Roy and A. Venugopalan, arXiv:quant-ph/9910004.
		%
		\bibitem{Fleming} C. H. Fleming, A. Roura and B. L. Hu, Ann. Phys. {\bf 326}, 1207 (2011).
		%
		\bibitem{SR} H. P. Robertson, Phys. Rev. {\bf 46}, 794 (1934).
		%
		\bibitem{Ref_5}
		H. Dekker and M. C. Valsakumar,
		Phys. Lett. {\bf 104A}, 67 (1984).
		%
		\bibitem{Ingarden} R. S. Ingarden and A. Kossakowski, Ann. Phys. {\bf 89}, 451 (1975).
		%
		\bibitem{Kraus} K. Kraus, {\it States, Effects, and Operations: Fundamental Notions of Quantum Theory} (Springer, Berlin, 1983).
		%
		\bibitem{Trifonov} D. A. Trifonov, Eur. Phys. J. B {\bf 29}, 349 (2002).
		%
		\bibitem{MOC} R. Courant and D. Hilbert, {\it Methods of Mathematical Physics} (Interscience Publishers, New York, 1962).
		%
		\bibitem{BCSH}
		J. Z. Bern{\'a}d, G. Homa and M. A. Csirik, 
		Eur. Phys. J. D {\bf 72}, 212 (2018).
		%
		\bibitem{KRGH} K. Robert and G. Hermann, Phys. Rev. E {\bf 55}, 153 (1997).
		%
		\bibitem{HFRR} H. Fritz and R. Reinhard, Phys. Rev. A {\bf 32}, 2462 (1985).
		%
		\bibitem{Hu_Paz} B. L. Hu, J. P. Paz and Y. Zhang, Phys. Rev. D {\bf 45}, 2843 (1992).	
		%
		\bibitem{yosida} K. Yosida, {\it Functional Analysis} (Springer-Verlag, Berlin, 1995).
		%
		\bibitem{Ref_11}
		V. Giovannetti and D. Vitali,
		Phys. Rev. A {\bf 63},  023812 (2001).
		
		
	\end{thebibliography}
\end{document}